\documentclass[journal]{IEEEtran}

\usepackage{comment}
\usepackage{url}
\usepackage{graphicx}
\usepackage{amsfonts}
\usepackage{stfloats}
\usepackage{soul}
\usepackage{amsmath}
\usepackage{epsfig}
\usepackage{multirow}
\usepackage{fancyhdr}
\usepackage{hhline}
\usepackage{url}
\usepackage{float}
\usepackage{enumitem}
\usepackage{cite}
\usepackage{multicol}
\usepackage{blindtext}
\usepackage{rotating}
\usepackage{tikz}
\usepackage{balance}
\usepackage{diagbox}
\usepackage{adjustbox}
\usepackage{rotating}
\usepackage{colortbl} 
\usepackage{array} 
\usepackage{xcolor} 
\usepackage{caption}
\usepackage{subcaption}

\usepackage[acronym,toc,shortcuts]{glossaries}

\newacronym{AI}{AI}{Artificial Intelligence}
\newacronym{AR}{AR}{Augmented Reality}
\newacronym{BQP}{BQP}{Bounded Error Quantum Polynomial Time}
\newacronym{CPU}{CPU}{Central Processing Unit}
\newacronym{CSI}{CSI}{Channel State Information}
\newacronym{DSA}{DSA}{Digital Signature Algorithm}
\newacronym{DoS}{DoS}{Denial of Service}
\newacronym{ECC}{ECC}{Elliptic Curve Cryptography}
\newacronym{eMBB}{eMBB}{enhanced Mobile Broadband}
\newacronym{FPGA}{FPGA}{Field-Programmable Gate Arrays}
\newacronym{FSO}{FSO}{Free Space Optical}
\newacronym{I/O}{I/O}{Input-Output}
\newacronym{IoT}{IoT}{Internet of Things}
\newacronym{KPI}{KPI}{Key Performance Indicator}
\newacronym{MEC}{MEC}{Mobile Edge Computing}
\newacronym{mMTC}{mMTC}{Massive Machine Type Communication}
\newacronym{MIMO}{MIMO}{Multiple-Input Multiple-Output}
\newacronym{ML-DSA}{ML-DSA}{Module-Lattice-Based Digital Signature Algorithm}
\newacronym{ML-KEM}{ML-KEM}{Module-Lattice-Based Key-Encapsulation Mechanism}
\newacronym{NISQ}{NISQ}{Noisy Intermediate-Scale Quantum}
\newacronym{NIST}{NIST}{National Institute of Standards and Technology}
\newacronym{QA}{QA}{Quantum Annealing}
\newacronym{QAOA}{QAOA}{Quantum Approximate Optimization Algorithm}
\newacronym{QBSA}{QBSA}{Quantum Blockchain Secured Algorithm}
\newacronym{QEC}{QEC}{Quantum Error Correction}
\newacronym{QGAN}{QGAN}{Quantum Generative Adversarial Network}
\newacronym{QC}{QC}{Quantum Computing}
\newacronym{PISQ}{PISQ}{Perfect Intermediate Scale Quantum Computing}
\newacronym{PLS}{PLS}{Physical Layer Security}
\newacronym{PQC}{PQC}{Post Quantum Cryptography}
\newacronym{QFT}{QFT}{Quantum Fourier Transform} 
\newacronym{QIP}{QIP}{Quantum Information Processing}
\newacronym{QKD}{QKD}{Quantum Key Distribution}
\newacronym{QML}{QML}{Quantum Machine Learning}
\newacronym{QNN}{QNN}{Quantum Neural Network}
\newacronym{QoS}{QoS}{Quality of Service}
\newacronym{QSDC}{QSDC}{Quantum Secure Direct Communication}
\newacronym{QPU}{QPU}{Quantum Processing Unit}
\newacronym{QWA}{QWA}{Quantum Walk Algorithm}
\newacronym{RAN}{RAN}{Radio Access Network}
\newacronym{RIS}{RIS}{Reconfigurable Intelligent Surfaces}
\newacronym{SDN}{SDN}{Software-Defined Networks}
\newacronym{SDK}{SDK}{Software Development Kit}
\newacronym{SLH-DSA}{SLH-DSA}{Stateless Hash-Based Digital Signature Algorithm}
\newacronym{SRPS}{SRPS}{Social Robots in Public Spaces}
\newacronym{RSA}{RSA}{Rivest–Shamir–Adleman}
\newacronym{THz}{THz}{Terahertz}
\newacronym{TRL}{TRL}{Technology Readiness Levels}
\newacronym{uRLLC}{uRLLC}{Ultra-Reliable and Low Latency Communication}
\newacronym{VAA}{VAA}{Virtual Antenna Array}
\newacronym{VAE}{VAE}{Variational Encoders}
\newacronym{VR}{VR}{Virtual Reality}
\newacronym{XR}{XR}{Extended Reality}

\begin{document}
\title{Quantum Technologies for Beyond 5G and 6G Networks: Applications, Opportunities, and Challenges}
    
\author
{Engin Zeydan,~\IEEEmembership{Senior Member,~IEEE}, Chamitha De Alwis,~\IEEEmembership{Senior Member,~IEEE}, Rabia Khan,~\IEEEmembership{Member,~IEEE},  Yekta Turk, Abdullah Aydeger~\IEEEmembership{Member,~IEEE}, ~Thippa~Reddy~Gadekallu, ~\IEEEmembership{Senior Member,~IEEE}, Madhusanka~Liyanage,~\IEEEmembership{Senior Member,~IEEE\vspace{-8mm}}

\thanks{Engin Zeydan is with Centre Tecnològic de Telecomunicacions de Catalunya (CTTC), Barcelona, Spain, 08860. e-mail: engin.zeydan@cttc.cat.}
\thanks{Chamitha De Alwis is with School of Computer Science and technology, University of Bedfordshire, United Kingdom. e-mail: chamitha@ieee.org.}
\thanks{Rabia Khan is with Scottish and Southern Electricity Networks (SSE) Transmission, Perth, Scotland email: rabia.khan@sse.com}
\thanks{Yekta Turk is with Aselsan Corp., Istanbul, Turkey, 34380. e-mail: yektaturk@aselsan.com.}
\thanks{Abdullah Aydeger is with the Department of Electrical Engineering and Computer Science at the Florida Institute of Technology, e-mail: aaydeger@fit.edu}
\thanks{Thippa Reddy Gadekallu is with The College of Mathematics and Computer Science, Zhejiang A\&F University, Hangzhou 311300, China, as well as with the Division of Research and Development, Lovely Professional University, Phagwara, India, and with the Center of Research Impact and Outcome, Chitkara University, Punjab, India, e-mail: thippareddy@ieee.org}
\thanks{Madhusanka Liyanage is with the School of Computer Science, University College Dublin, Ireland, email: madhusanka@ucd.ie }
}

\maketitle


\begin{abstract}




As the world prepares for the advent of  6G networks, quantum technologies are becoming critical enablers of the next generation of communication systems. This survey paper investigates the convergence of quantum technologies and 6G networks, focusing on their applications, opportunities and challenges. We begin with an examination of the motivations for integrating quantum technologies into 6G, investigating the potential to overcome the limits of classical computing and cryptography. We then highlight key research gaps, particularly in quantum communication, quantum computing integration and security enhancement.  A comprehensive overview of quantum technologies relevant to 6G, including quantum communication devices, quantum computing paradigms, and hybrid quantum-classical approaches is provided. A particular focus is on the role of quantum technologies in enhancing 6G Radio Access Networks (RAN),  6G core and edge network optimization, and 6G security. The survey paper also explores the application of quantum cryptography with a focus on Quantum Key Distribution (QKD), Quantum Secure Direct Communication (QSDC) and quantum-resistant cryptographic algorithms and assesses their implementation challenges and potential impact on 6G networks. We also discuss the significant challenges associated with integrating quantum technologies into existing communications infrastructures, including issues of technological maturity, standardization, and economic considerations. Finally, we summarize the lessons learned from current research and outline future research directions to guide the ongoing development of quantum-enabled 6G networks.
\end{abstract}



\section{Introduction} \label{intro} 



The global rollout of fifth-generation (5G) wireless communication networks began in 2020, introducing various standardized features such as mass connectivity, ultra- reliability and low latency \cite{bovzanic2021mobile}. Despite these advances, 5G is not expected to fully meet future requirements beyond 2030. Sixth generation (6G) wireless communication networks are therefore expected to emerge, offering global coverage, improved spectrum/cost and energy efficiency, greater security and intelligence are some of their benefits \cite{chowdhury20206g}. \textcolor{black}{However,} the transition from 5G to 6G networks is a big leap in the evolution of wireless communications, promising unprecedented levels of performance, reliability and security. As we move towards the introduction of 6G, it is crucial to leverage new technologies that can meet these increased expectations. 6G networks will rely on cutting-edge technologies. One of these promising possibilities is quantum technology \cite{dowling2003quantum}. The main reasons for integrating quantum technologies into 6G networks will be to meet the increasing service and user demands \cite{wang2022quantum, rozenman2023quantum, alshaer2024ai}.

\begin{figure}[htp!]
    \centering
    \includegraphics[width=0.7\linewidth]{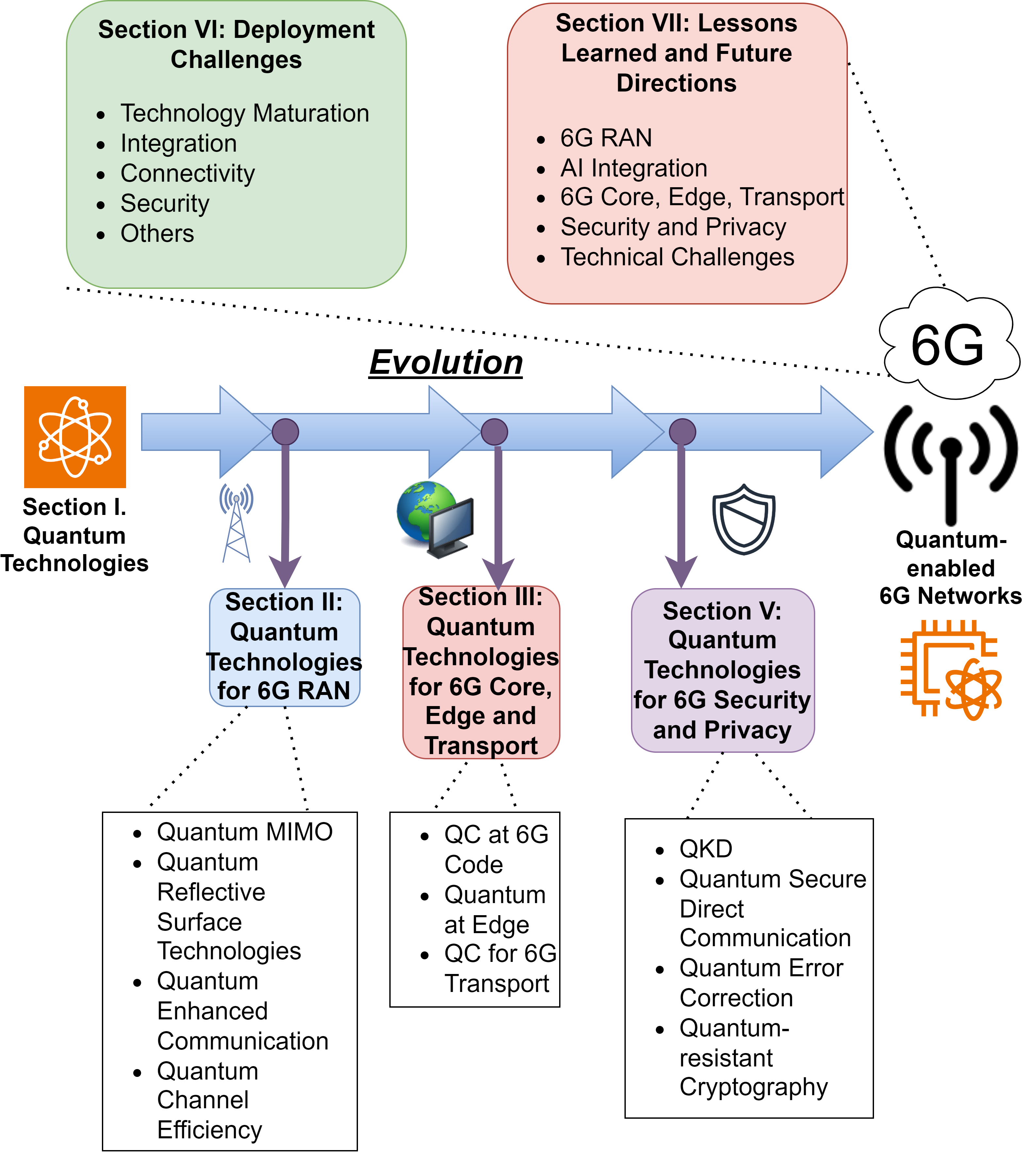}
    \caption{Quantum-enabled 6G vision projected in the paper along with the paper structure.}
    \label{fig:Quantum_6G}
    \vspace{-.2cm}
\end{figure}

\textcolor{black}{Classical cryptographic methods—such as RSA, DSA, and ECC—form the foundation of security in current communication systems, including 5G networks~\cite{schinianakis2017alternative}. These methods depend on the computational intractability of mathematical problems like integer factorization and discrete logarithms. However, the emergence of quantum computing poses a serious threat to these systems. Algorithms such as Shor’s algorithm can efficiently factor large integers, rendering RSA and related schemes vulnerable to quantum attacks~\cite{shor1994algorithms, nandhini2022extensive}. In parallel, quantum computing is rapidly advancing from a theoretical concept to a practical technological discipline. The field of Quantum Information Processing (QIP) focuses on the manipulation of quantum data using principles such as entanglement and superposition. These properties enable exponential parallelism, which in turn promises groundbreaking improvements in computational speed, energy efficiency, and security~\cite{de2021survey}. This shift is reflected in global market projections, which estimate the quantum computing industry will reach USD 143.44 billion by 2032, growing at a CAGR of 26.5
The evolution toward 6G networks is driven by a need for ultra-fast, intelligent, and secure communication systems. 6G is expected to outperform 5G in areas such as latency, data throughput, device density, AI-native architecture, and end-to-end security, supporting futuristic applications like real-time holography and quantum cloud computing (see Table~\ref{6G_5G_comparison}). However, most 6G-enabling technologies—including quantum components—remain in early developmental phases, typically between TRL 1 and 3. Figure~\ref{fig:Quantum_6G} illustrates how quantum technologies are envisioned to evolve toward integration within future 6G networks.}

\begin{table*}[htp!]
\captionsetup{font=sc}
\centering
\scriptsize
\caption{Detailed Comparison of 5G and 6G Capabilities with TRL Stages}
\label{6G_5G_comparison}
\begin{tabular}{|p{1.5cm}|p{3.5cm}|p{5cm}|p{5cm}|p{1cm}|}
\hline
\rowcolor{gray!30}
\textbf{Feature}               & \textbf{5G}                          & \textbf{6G (Projected)}                         & \textbf{Key Differences/Advantages (6G over 5G)} & \textbf{TRL for 6G} \\ \hline
\textbf{Data Rate}             & Up to 20 Gbps \cite{shafi20175g}  & Multi-terabit-per-second speeds \cite{dang2020should} & 10-100x increase in speed, enabling real-time holography & [2-3]  \\ \hline
\textbf{Latency}               & 1 ms \cite{parvez2018survey}          & Target 100 µs \cite{dang2020should}    & Reduced latency allows tactile Internet, real-time AI      & [2]  \\ \hline
\textbf{Coverage}              & Urban and some rural areas & Global, including remote, space, and underwater & Expanded access with satellite-terrestrial integration     & [1-2]  \\ \hline
\textbf{Network Density}       & 1 million devices/sq km \cite{shafi20175g} & 10 million devices/sq km \cite{dang2020should} & Greater device interconnectivity for IoT, smart cities     & [2-3]  \\ \hline
\textbf{AI Integration}        & AI-driven network management  & Quantum-enhanced AI for autonomous operation  & Near-zero human intervention with self-optimizing networks & [1-2]  \\ \hline
\textbf{Security}              & Classical cryptography  & Quantum-secured communication (QKD, PQC) \cite{dang2020should} & Stronger against quantum attacks, more scalable           & [1-2]  \\ \hline
\textbf{Use Cases}             & IoT, AR/VR, autonomous vehicles  & Real-time holography, intelligent transportation, quantum cloud computing  & More advanced applications like telepresence, quantum AI   & [1-2]  \\ \hline
\end{tabular}
\vspace{-5mm}
\end{table*}

\subsection{Motivation and Contribution}

Despite the potential benefits, several research gaps still need to be filled in order to fully integrate quantum technologies into B5G and 6G networks. \textcolor{black}{These gaps are rooted in the fundamental concepts outlined in Section \ref{sec:background}, where  the basics of quantum technologies, quantum threats, and hybrid quantum-classical systems are introduced. Specifically:}  \textit{(i)} The practical implementation of quantum communication protocols in real networks remains a major challenge. \textcolor{black}{As highlighted in Section \ref{subsec:quantumtech} and \ref{subsec:quantumthreats}, issues such as} quantum channel noise, error correction and maintaining entanglement over long distances still \textcolor{black}{require significant advances in both algorithms and hardware.} 
\textit{(ii)}  The integration of quantum computing with AI for network management and optimization, \textcolor{black}{as discussed in Section \ref{subsec:hybrid} and later explored in depth in Section \ref{sec:quantum} calls for} the development of new algorithms and frameworks that can take advantage of quantum computing.  \textit{(iii)}  The development of scalable quantum technologies that can be used on a commercial scale is crucial. \textcolor{black}{Building on the discussion in Section \ref{subsec:products},} this includes the development of quantum hardware that is cost-efficient, reliable and compatible with existing network infrastructures. \textit{(iv)} The establishment of standards and protocols for the use of quantum technologies in communication networks is crucial to ensure interoperability and broad acceptance.

The main motivation for exploring quantum technologies in the context of B5G and 6G networks is the growing need for improved communication capabilities that transcend the limits of classical systems. The exponential growth of connected devices, the demand for real-time data processing and the need for robust security measures are driving the trend towards more advanced networking solutions. Quantum technologies, with their potential for high-speed computation, secure communication and efficient data processing, offer a promising way to achieve these goals. Quantum computers can perform complex calculations at speeds that are unattainable for classical computers. This makes them ideal for tasks such as network optimization and AI training. Quantum communication technologies such as \ac{QKD} offer theoretically unbreakable encryption and increase the security of data transmission in 6G networks. Quantum algorithms can process large data sets more efficiently, improving the overall performance of AI and machine learning applications in network management. This paper aims to address above research gaps by providing a detailed examination of how quantum technologies can be applied to B5G and 6G networks. The key contributions of this paper are:

\begin{itemize}
    \item We provide a comprehensive overview of how quantum technologies in particular advancements in \ac{QC} can enhance 6G network capabilities, especially in areas of 6G core, edge, \ac{RAN}, transport. \textcolor{black}{These contributions build directly on the foundational concepts introduced in Section \ref{sec:background} and extend them with domain-specific applications.} 
    \item We highlight quantum-based security mechanisms like \ac{QKD}, quantum-secure direct communication, and quantum-resistant cryptography to mitigate evolving security threats in 6G. These technologies promise enhanced protection against quantum computing attacks and secure data transmission. \textcolor{black}{These are grounded in the quantum threat models discussed in Section \ref{subsec:quantumthreats} and the technologies explored in Section \ref{sec:security}.}
    \item We identify the key technical, connectivity and integration challenges in deploying quantum technologies in the 6G core, edge and transport layers. We also discuss potential solutions such as hybrid quantum-classical systems, quantum error correction techniques and energy-efficient quantum processors to ensure scalable and efficient integration, \textcolor{black}{reflecting on the challenges introduced in Section \ref{sec:deployment}}.
    \item We outline lessons learned and future areas of research, including quantum-enabled transport protocols, quantum-resistant cryptographic frameworks, and error correction methods for quantum-enabled communication systems. We also present ongoing quantum projects and standardization efforts to support the development of quantum technologies for 6G.
\end{itemize}

Finally, Table \ref{tab:summaryofprojects} summarizes important surveys related to quantum technology and shows how our work differs from them \textcolor{black}{ in terms of scope, depth, and future-oriented synthesis}.

\begin{table*}
 \captionsetup{font=sc,  position=above, justification=centering, labelsep=newline,singlelinecheck=true}
 \scriptsize
  \centering
        \caption{Summary of Important Surveys on Quantum Communication and Technology}
        \renewcommand{\arraystretch}{1.2}
  \begin{tabular}{|p{1.5cm}||c|c|c|c|c|p{10.5cm}|}  
  \hline 
   \rowcolor{gray!25}      	
         &{\rotatebox[origin=c]{90}{\begin{tabular}[c]{@{}c@{}}Quantum technologies  \\ for 6G RAN\end{tabular} }}
         &{\rotatebox[origin=c]{90}{\begin{tabular}[c]{@{}c@{}}Quantum technologies for \\ 6G Core and Edge\end{tabular} }}
         &{\rotatebox[origin=c]{90}{\begin{tabular}[c]{@{}c@{}}Quantum technologies for \\ 6G Security and Privacy\end{tabular} }}
         &{\rotatebox[origin=c]{90}{\begin{tabular}[c]{@{}c@{}}Deployment Challenges\end{tabular} }}
         &{\rotatebox[origin=c]{90}{\begin{tabular}[c]{@{}c@{}}Future directions analysis \\ and lessons learned on \\ Quantum in 6G\end{tabular} }}         
         & Remarks \\
  \hline
\hline
Prateek et al. 2023 \cite{prateek2023quantum} & \cellcolor{red!15}L & \cellcolor{red!15}L & \cellcolor{green!15}H & \cellcolor{yellow!15}M & \cellcolor{green!15}H & Discusses the application of quantum security technologies, including QKD and post-quantum cryptography, in securing 6G-enabled IoE environments, highlighting the challenges and potential solutions for large-scale deployment. \\

\hline
Glisic et al. 2024 \cite{glisic2024quantum} & \cellcolor{red!15}L & \cellcolor{red!15}L & \cellcolor{green!15}H & \cellcolor{red!15}L & \cellcolor{green!15}H & Provides a detailed comparison between quantum security mechanisms and post-quantum cryptography, analyzing their suitability and potential integration into future network architectures, including B5G/6G. \\

\hline
Zhou et al. 2023 \cite{zhou2023survey} & \cellcolor{red!15}L & \cellcolor{red!15}L & \cellcolor{green!15}H & \cellcolor{red!15}L & \cellcolor{green!15}H & Analyzes the readiness and challenges of implementing post-quantum cryptographic algorithms in 5G/6G networks, focusing on their security, performance, and integration with existing protocols. \\

\hline
Javeed et al. 2024 \cite{javeed2024quantum} & \cellcolor{red!15}L & \cellcolor{green!15}H & \cellcolor{green!15}H & \cellcolor{red!15}L & \cellcolor{green!15}H & Explores the integration of quantum computing with federated learning for enhancing IoT security in 6G networks, identifying key challenges and proposing future research directions. \\

\hline
Jawad et al. 2023 \cite{jawad2023comprehensive} & \cellcolor{red!15}L & \cellcolor{red!15}L & \cellcolor{yellow!15}M & \cellcolor{yellow!15}M & \cellcolor{yellow!15}M & Provides an extensive overview of enabling technologies for 6G, including the role of machine learning in optimizing network performance, along with the challenges and opportunities these technologies present. \\

\hline
Kumar et al. 2021 \cite{kumar2021survey} & \cellcolor{yellow!15}M & \cellcolor{yellow!15}M & \cellcolor{yellow!15}M & \cellcolor{red!15}L & \cellcolor{yellow!15}M & Discusses the potential of quantum technologies in enhancing the capabilities of drones and their networks, focusing on communication, navigation, and security aspects, with implications for future 6G applications. \\

\hline
Xiao et al. 2024 \cite{xiao2024securing} & \cellcolor{red!15}L & \cellcolor{red!15}L & \cellcolor{green!15}H & \cellcolor{red!15}L & \cellcolor{green!15}H & Examines the role of physical-layer key generation in enhancing the security of NextG networks, focusing on the techniques, challenges, and practical applications in the context of 6G. \\

\hline
Al et al. 2024 \cite{al2024characterizing} & \cellcolor{yellow!15}M & \cellcolor{yellow!15}M & \cellcolor{yellow!15}M & \cellcolor{red!15}L & \cellcolor{green!15}H & Explores the synergies between quantum technologies and non-terrestrial networks, such as satellite communications, highlighting the potential for enhanced security and performance in 6G and beyond. \\

\hline
Bhat et al. 2021 \cite{bhat20216g} & \cellcolor{yellow!15}M & \cellcolor{red!15}L & \cellcolor{yellow!15}M & \cellcolor{yellow!15}M & \cellcolor{yellow!15}M & Provides a detailed analysis of the current status of 6G development, explores future perspectives, and discusses key enabling technologies, challenges, and use cases for 6G networks. \\

\hline
Bashirpour et al. 2022 \cite{bashirpour2022quantum} & \cellcolor{red!15}L & \cellcolor{yellow!15}M & \cellcolor{green!15}H & \cellcolor{yellow!15}M & \cellcolor{green!15}H & Analyzes the potential of quantum technologies to enhance the security, efficiency, and sustainability of smart cities, focusing on communication networks, data processing, and infrastructure management within the context of 6G. \\

\hline
Bouchmal et al. 2023 \cite{bouchmal2023classical} & \cellcolor{yellow!15}M & \cellcolor{yellow!15}M & \cellcolor{yellow!15}M & \cellcolor{red!15}L & \cellcolor{green!15}H & Explores the transition from classical to quantum machine learning for optimizing routing in 6G \ac{SDN}, highlighting the advantages, challenges, and potential future directions of quantum-enhanced approaches. \\

\hline
Duong et al. 2022 \cite{duong2022quantum} & \cellcolor{red!15}L & \cellcolor{red!15}L & \cellcolor{green!15}H & \cellcolor{yellow!15}M & \cellcolor{green!15}H & Discusses the fundamentals of quantum-inspired machine learning, its applications in 6G for security and resource allocation, and the challenges and future research directions in integrating these technologies into next-generation networks. \\

\hline
Narottama et al. 2023 \cite{narottama2023quantum} & \cellcolor{yellow!15}M & \cellcolor{yellow!15}M & \cellcolor{yellow!15}M & \cellcolor{red!15}L & \cellcolor{green!15}H & Explores the fundamental concepts of quantum machine learning and its potential applications in next-generation wireless communication systems, including 6G, with an emphasis on challenges and future research directions. \\

\hline
Glisic et al. 2024 \cite{glisic2024quantum} & \cellcolor{yellow!15}M & \cellcolor{yellow!15}M & \cellcolor{green!15}H & \cellcolor{yellow!15}M & \cellcolor{green!15}H & Investigates the intersection of quantum computing and neuroscience in the context of 6G/7G networks, discussing the potential for enhancing network intelligence, security, and efficiency, and identifying key challenges and future research avenues. \\

\hline
Zhang et al. 2022 \cite{zhang2022future} & \cellcolor{yellow!15}M & \cellcolor{yellow!15}M & \cellcolor{green!15}H & \cellcolor{red!15}L & \cellcolor{green!15}H & Provides a comprehensive analysis of quantum-enhanced distributed learning systems for 6G networks, focusing on their potential applications, challenges, and proposing future research directions for integrating quantum computing with AI-driven 6G systems. \\

\hline
Liu et al. 2022 \cite{liu2022towards} & \cellcolor{red!15}L & \cellcolor{red!15}L & \cellcolor{green!15}H  & \cellcolor{red!15}L & \cellcolor{yellow!15}M & Examines the challenges and opportunities for the large-scale deployment of QKD in communication networks, focusing on technological, economic, and regulatory aspects necessary for industrialization. \\

\hline
Singh et al. 2021 \cite{singh2021quantum} & \cellcolor{yellow!15}M & \cellcolor{yellow!15}M & \cellcolor{green!15}H & \cellcolor{red!15}L & \cellcolor{yellow!15}M & Provides an in-depth analysis of the quantum internet, including its potential applications, the enabling technologies required, the challenges faced, and the future research directions needed to realize its full potential. \\

\hline
Tang et al. 2023 \cite{yang2023survey} & \cellcolor{yellow!15}M & \cellcolor{yellow!15}M & \cellcolor{green!15}H & \cellcolor{red!15}L & \cellcolor{yellow!15}M & Discusses the critical issues in integrating quantum computing and communications, including scalability, error correction, and network security, and provides insights into possible solutions and future research directions. However, the focus is not primarily on B5G and 6G. \\

\hline
Mousa et al. 2024 \cite{mousa2024survey} & \cellcolor{red!15}L & \cellcolor{red!15}L & \cellcolor{yellow!15}M & \cellcolor{red!15}L  & \cellcolor{yellow!15}M & This paper provides an analysis of quantum computing adoption with a focus on privacy engineering, discussing challenges and considerations for integrating quantum computing into privacy frameworks. While it does not directly address 6G, its insights on privacy and security are relevant to future network technologies. \\

\hline
This survey & \cellcolor{green!15}H & \cellcolor{green!15}H & \cellcolor{green!15}H & \cellcolor{green!15}H & \cellcolor{green!15}H & This survey explores the applications, opportunities, and challenges of integrating quantum technologies into B5G/6G networks, covering key areas such as quantum communication for 6G RAN, core, edge, transport, security and privacy. Identifies research gaps and future directions to guide ongoing development in this field. \\

\hline

  \end{tabular}
  \begin{flushleft}
\begin{center}
\begin{tikzpicture}
\node (rect) at (-7,2) [draw,thick,minimum width=0.6cm,minimum height=0.4cm, fill= green!15, label=0:Explores the field in detail.] {H};
\node (rect) at (-2.5,2) [draw,thick,minimum width=0.6cm,minimum height=0.4cm, fill= yellow!15, label=0:Provides some information about the field.] {M};
\node (rect) at (4,2) [draw,thick,minimum width=0.6cm,minimum height=0.4cm, fill= red!15, label=0:No information or explores the area only briefly.] {L};
\end{tikzpicture}
\end{center}
\end{flushleft}
\label{tab:summaryofprojects}
\vspace{-0.5cm}
\end{table*}

\subsection{Outline}
 
To provide a clear and structured discussion, the paper is organized \textcolor{black}{to mirror the functional architecture of a 6G communication system. Unlike previous survey works that treat quantum technologies in isolation, this paper adopts a **layered and system-oriented structure**, mapping quantum capabilities directly to the **radio access, core, edge, transport, AI, and security layers** of 6G networks. This layered approach offers both a comprehensive overview and practical insights into how quantum technologies can be integrated across the 6G stack. The structure is summarized} in Fig. \ref{fig:Quantum_6G}: Section \ref{sec:background} introduces the fundamental concepts of quantum technologies and 6G networks, \textcolor{black}{covering quantum computing, quantum communication, quantum threats, and hybrid classical-quantum paradigms.}

Section \ref{sec:quantum} explores how quantum technologies can be applied to \ac{RAN} in 6G, including specific implementations at the physical layer \textcolor{black}{such as quantum MIMO, reflective surface technologies} and their potential impacts.  Section \ref{sec:core} discusses \textcolor{black}{how quantum computing and quantum-assisted optimization can enhance the 6G core, edge, and transport domains,} highlighting specific use cases and implementations, \textcolor{black}{with emphasis on routing, traffic management, and latency reduction}.  Section \ref{sec:security} \textcolor{black}{explores} the security enhancements provided by quantum technologies, including QKD, \ac{QSDC}, and \textcolor{black}{quantum error correction, post-quantum cryptography and} other advanced security mechanisms.  Section \ref{sec:lesson} identifies and analyzes the major challenges in deploying quantum technologies within B5G and 6G networks, providing insights into possible solutions and strategies for overcoming these challenges. Section \ref{sec:lesson} summarizes the lessons learned from current research, identifies remaining research questions, and outlines potential future research directions to advance the integration of quantum technologies in communication networks. Section \ref{sec:conclusion} provides the conclusions of the paper.


\section{Background} 
\label{sec:background}

In this section we discuss in brief about the basic fundamentals of QC, 6G, and how the integration of QC is beneficial for 6G.




\subsection{Quantum Technologies}
\label{subsec:quantumtech}

QC is based on the principles of fundamental physics. The basic unit of any physical property is known as quantum in physics. The term quantum in QC refers to the system that is used to calculate the outputs in quantum mechanics. Quantum is based on the properties of the basic atomic or subatomic particles of any physical entity such as photons, neutrons, and electrons \cite{yang2023survey,sood2023quantum}.

\subsubsection{Key Terminologies in QC}
Some of the key terminologies in QC are as follows: \textit{(i) Qubit} is an elementary unit of information in QC. Qubits can store all the possible combinations of zeroes and ones. \textit{(ii) Superposition}: Superposition of the information in qubit refer to all the possible permutations and combinations of the qubit. Superposition of qubits can help in the creation of multidimensional computational spaces, through which the representation of complex problems in new ways is possible.  \textit{(iii) Entanglement}: It is an effect in which the behavior of two separate things is correlated. Changes in qubit affect another when two qubits are entangled. \textit{(iv) Interference}: When the entangled qubits are superpositioned. These numerous probabilities are build on each other when many of the probabilities peak at a particular result or when they cancel each other when the troughs and peaks interact with each other. Both of these are forms of interference.

\subsubsection{Classical Computing vs QC}

Today's smartphones, laptops, supercomputers and desktops are based on classical computing, which is built on bits. Each bit can store one piece of information, which can be either a zero or a one. QC, on the other hand, is based on quantum bits, also known as "qubits", which can store 0s or 1s or any combination of 0s and 1s simultaneously. The chips in QC are the physical storage that can store qubits, similar to classical computing. In classical computing, when a problem with multiple variables needs to be solved, a computer must perform a calculation for each change in the value of a variable \cite{pan2024evolution}. Every calculation in classical computing is a single path that leads to a single result. In QC, however, many paths can be explored for each computation. In addition, QC qubits can interact with each other, which enables exponential scalability of the qubits, so that QC can solve extremely complex problems very quickly compared to classical computing \cite{chen2024nisq}.

Many industries such as global shipping companies, financial institutions, the defense sector, engineering firms and the communications industry can use \ac{QC} to solve important and complex problems in their fields \cite{corcoles2019challenges}. Table \ref{tab:characterists_quantum} contains a detailed comparison between classical and quantum computing for network applications.

\begin{table*}[htp!]
 \captionsetup{font=sc,  position=above, justification=centering, labelsep=newline,singlelinecheck=true}
\centering
\scriptsize
\caption{Detailed Comparison of Classical vs. Quantum Computing for Network Applications}
\label{tab:classical_quantum}
\begin{tabular}{|p{1.5cm}|p{2.5cm}|p{4cm}|p{8cm}|}
\hline
\rowcolor{gray!30}
\textbf{Feature}               & \textbf{Classical Computing}         & \textbf{Quantum Computing}                      & \textbf{Application Impact in 6G Networks}       \\ \hline
\textbf{Basic Unit}            & Bit (0 or 1)                        & Qubit (0, 1, or both simultaneously)            & The data can be transmitted in exponentially faster time due to the superior problem-solving capabilities of QC \cite{mohammed2023quantum} \\ \hline
\textbf{Processing Speed}       & clock speed measured in gigahertz                      & Clock Speed measured in teraflops           & Solves complex routing and optimization problems in 6G networks faster \cite{akbar2024role} \\ \hline
\textbf{Data Security}         & Vulnerable to quantum attacks       & Quantum-resistant encryption (QKD, PQC)         & Provides unbreakable encryption with QKD \cite{prateek2023quantum}        \\ \hline
\textbf{Hardware Complexity}   & Well-established and cost-effective & Resource-intensive and expensive                & Quantum systems are complex, but they are more suitable for performing critical tasks in 6G such as resource optimization, and data processing complexity \cite{rozenman2023quantum} \\ \hline
\textbf{Energy Efficiency}     & High energy consumption             & Lower energy consumption for complex calculations & QC is more efficient for large-scale AI and network optimizations in 6G \cite{urgelles2024network} \\ \hline
\textbf{Error Handling}        & Error-prone, needs redundancy       & QEC can identify and fix errors in quantum computers in quick time           & QEC can help reliable and fault tolerant communication over noisy channels in 6G \cite{krinner2022realizing} \\ \hline
\textbf{Use Case}              & Traditional AI and network management & AI, cryptography, optimization in 6G             & Quantum systems can handle tasks unfeasible for classical systems \cite{duong2022quantum}\\ \hline
\end{tabular}
\end{table*}

\subsection{QC for B5G and 6G Networks}
\label{subsec:hybrid}

\textcolor{black}{ 
6G is envisioned to significantly surpass current 5G capabilities, enabling a wide array of new use cases and applications. Compared to 5G, 6G aims to drastically improve data transmission rates, latency, and connection density, supporting advanced services in areas such as imaging, artificial intelligence, immersive communication, and healthcare. However, to fully realize the transformative potential of 6G, several technical and infrastructural challenges must be addressed. These challenges include the need for power-efficient processing units, ultra-high-frequency transceivers, and advanced network optimization techniques \cite{dang2020should}.
}

\textcolor{black}{\ac{QC} has emerged as a promising enabler to overcome many of these challenges and accelerate the deployment and success of 6G networks \cite{akbar2024role}. Next, we outline several key areas where quantum technologies can contribute to 6G:
(i) Advanced data processing: QC offers the ability to process massive amounts of data simultaneously. This capability can significantly enhance the data processing capacity of 6G, improving network speed, traffic management, reliability, and energy efficiency \cite{dang2020should, Farouk2024quantum}.
(ii) \ac{QKD}: Security is a fundamental concern in next-generation communication systems. QKD utilizes principles of quantum mechanics to securely distribute cryptographic keys, offering theoretically unbreakable security. Integrating QKD can add a robust layer of protection to the massive IoT ecosystems within 6G \cite{Wang2020security}.
(iii) \ac{QML}: Machine learning plays a critical role in optimizing and maintaining 6G networks. QML can further enhance this by enabling faster training and more accurate predictive models, leading to intelligent and self-optimizing networks \cite{Narottama2024quantum}.
(iv) Quantum sensors: Precision in positioning, navigation, and timing is crucial in 6G applications. Quantum sensors can provide extremely high-resolution measurements, enhancing localization accuracy in devices and boosting network context-awareness \cite{Bongs2023quantum}.
(v) Quantum clocks: Precise time synchronization is essential in applications like industrial automation and autonomous driving. Quantum clocks offer ultra-high precision, improving synchronization across 6G nodes and reducing latency \cite{Nande2023quantum}.
(vi) Resource and interference management: With the increased heterogeneity and density of devices in 6G, resource allocation and interference mitigation become significantly more complex. QC can tackle these large-scale optimization problems more efficiently than classical methods \cite{zaman2023quantum, Aliaga2024modeling, Farouk2024quantum}.
(vii) Quantum error correction: Reliable data transmission is key to sustaining high-performance communication networks. Quantum error correction techniques can protect quantum information against noise and improve the robustness of 6G systems \cite{Qu2023QEPP}.
(viii) Signal processing: QC can reduce latency and increase throughput by performing complex signal processing tasks more efficiently. This is critical for enhancing the overall quality of service in 6G \cite{Farouk2024quantum}.
(ix) Network design and planning: Optimizing network infrastructure—such as base station placement, routing paths, and spectrum usage—requires solving complex combinatorial problems. QC can assist in making more efficient decisions during the planning and design phases of 6G networks \cite{Phillipson2023quantum}.
(x) Quantum computing at the edge: Edge computing is central to low-latency and context-aware services in 6G. Deploying quantum computing capabilities at the network edge can support real-time analytics, autonomous operations, and secure localized processing \cite{Qu2023QEPP}.
(xi) Low-power device optimization: 6G will connect billions of low-power IoT devices in applications like smart cities and homes. QC can enable energy-efficient computations and resource management, enhancing the viability of such large-scale, power-constrained deployments \cite{kamruzzaman2022key}.
}

Quantum enabled 6G is illustrated in Fig. \ref{fig:Quantum-6G}. The physical channels will be connected by quantum capable mediums such as fiber-optic or laser cable. Some of the nodes will also be having fully quantum capability, but most of the nodes can support generation, transmission and reception of qubits. The layers in the quantum enabled 6G networks are summarized briefly below \cite{wang2022quantum}:

\begin{figure*}[htp!]
    \centering
     \includegraphics[width=\linewidth]{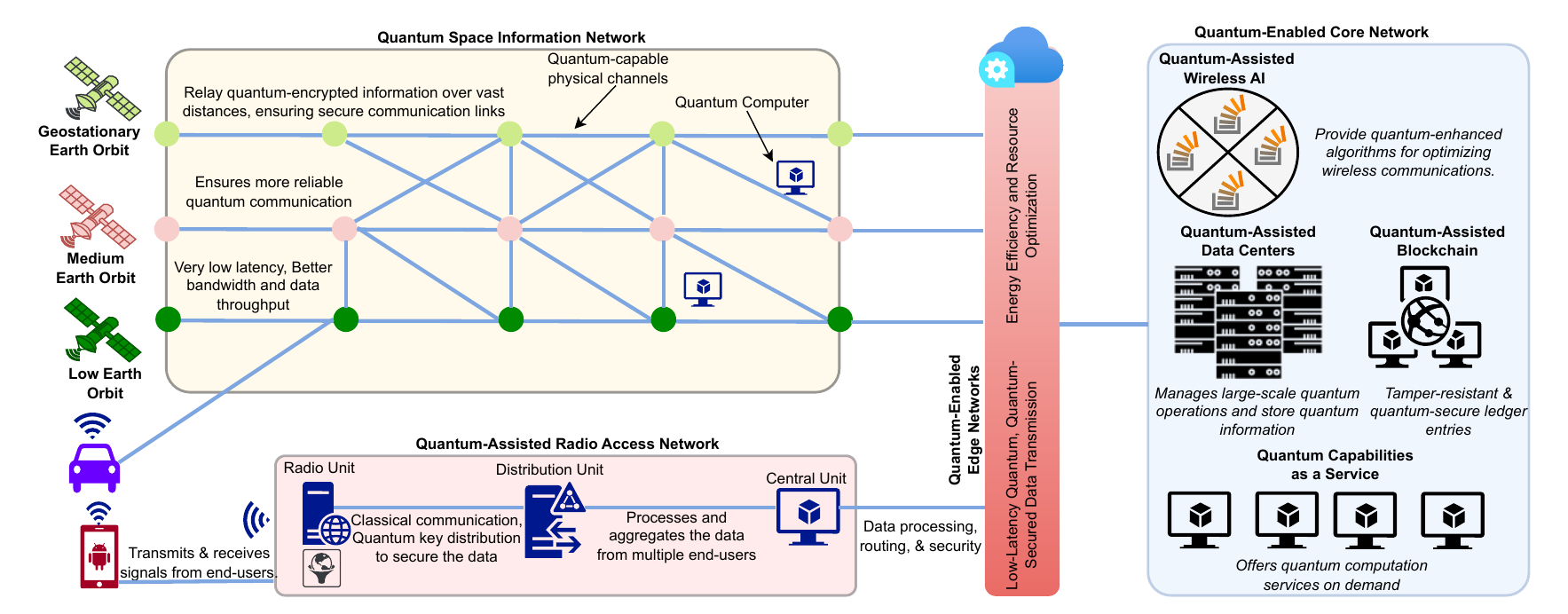}
    \caption{Quantum enabled 6G illustrated with corresponding layers.}
    \label{fig:Quantum-6G}
    \vspace{-.5cm}
\end{figure*}

\subsubsection{Quantum-Assisted Radio Access Networks (qRAN)} QC technologies and algorithms can improve security and efficiency of  \ac{RAN} \cite{arabul2023100}. By adapting algorithms such as quantum search algorithms, finding the optimal solution for cells planning and allocation of radio access, spectrum and energy efficiency can be achieved in 6G \cite{al2024optimizing}. Quantum technologies such as QKD can secure O- \ac{RAN} components by establishing secured keys among them \cite{gur20236g}.

\subsubsection{Quantum Space Information Networks} With the advancement of 6G, numerous satellites are expected to be deployed in several space orbits to form space information networks. Free-space optic links are used to connect the satellite nodes and a quantum computer can be hosted on a powerful satellite node. Quantum space information network (QSIN) can enhance the quantum communications ground stations and a satellite node. Moreover, QSIN can provide quantum computing as a service for ground stations and other satellite nodes. Also, QSIN can act as a quantum repeater or a trust node to improve the quantum communications \cite{bas2023review, duong2023machine}.     

\subsubsection{Quantum-Assisted Edge Networks} Edge computing is expected to play a major role in realizing the full potential of 6G. Hence, numerous edge nodes are expected to be connected to the 6G. Some of the challenges the edge computing are resource allocation in edge, task offloading, and security. QC can help in addressing these issues by finding optimal solutions for edge resource allocation and task offloading, while secure quantum communications can be used to secure the communication between the edge nodes \cite{ansere2023quantum, garcia2023secure}. 

\subsubsection{Quantum-Assisted Data Center} Modern data centers use optical links like free-space optics and optical fiber to reduce interference and yo increase data rates for inter-rack communications \cite{kazemi2023terabit}. To improve inter-rack security, quantum cryptography and quantum communications can be realized by leveraging quantum channels in these optical links. QC can also help on solving complex problems in data centers like management of energy consumption and optimal data flow management \cite{zaman2023quantum}.

\subsubsection{Quantum-Assisted Blockchain} Blockchain is a key enabler of 6G, which can be used in several applications like distributed wireless resource sharing and decentralized authentication in parties that don't trust each other. However some issues such as privacy breaches from the transparent data in the nodes of blockchain, low transmission speeds due to consensus algorithms, and security risks from malicious nodes,  are some of the issues of blockchain technology. Quantum assisted blockchain can solve these issues. For instance, the security in communications among the nodes in blockchain nodes can be improved by quantum communication algorithms; link relation between the blocks in the blockchain can be simulated by non-separability of entangled qubits; new consensus protocols can be designed by entanglement without inducing communication overheads, that can lead to increased transaction speeds \cite{sharma2023qiotchain}.

\subsubsection{Quantum-Assisted Wireless Artificial Intelligence} Application of artificial intelligence algorithms such as transfer learning, federated learning, deep learning, reinforcement learning, etc will make 6G networks more autonomous and intelligent. QC can help 6g in several ways. Federated learning, which is a decentralized AI technology, introduce model exchange and communications between the participants, that can be secured by quantum cryptography \cite{zhang2022federated}. QC can help in improving the processing speeds of interference process and training process in AI. More efficient and robust wireless AI algorithms can be introduced by
    quantum machine learning \cite{narottama2023quantum}.

\textcolor{black}{Table~\ref{tab:quantum_algorithms} summarizes key quantum algorithms and techniques relevant to communication and networking. It provides a quick reference for foundational concepts in quantum computing and quantum communication, linking each algorithm to its purpose and potential applications in next-generation systems such as 6G networks.  }



\begin{table*}[htp!]
\centering
\color{black}
\caption{Summary of Quantum Algorithms for Communication and Network Optimization}
\label{tab:quantum_algorithms}
\scriptsize
\begin{tabular}{|p{3.2cm}|p{4.5cm}|p{4cm}|p{4cm}|}
\hline
\rowcolor{gray!30}
\textbf{Algorithm/Technique} & \textbf{Description} & \textbf{Purpose} & \textbf{Applications in 6G/Networks} \\ \hline

\textbf{BB84 Protocol (QKD)} & Uses quantum states (e.g., photon polarization) to generate shared secret keys & Ensure secure communication & Encrypted communication in 6G, sensitive data transfer \\ \hline

\textbf{Superdense Coding} & Sends 2 classical bits using 1 qubit and entanglement & Boost classical channel capacity & High-efficiency data exchange in limited-bandwidth settings \\ \hline

\textbf{Quantum Teleportation} & Transfers quantum state using entangled qubits and classical communication & Reliable quantum state transfer over distance & Quantum link layer in quantum networks \\ \hline

\textbf{Quantum Repeaters} & Use entanglement swapping and purification to extend transmission range & Long-distance quantum communication & Quantum internet, secure satellite communication \\ \hline

\textbf{E91 Protocol (Entanglement-based QKD)} & Uses entangled particles and Bell's theorem for secure key exchange & Eavesdropping detection & Next-gen QKD systems in 6G \\ \hline

\textbf{Quantum Error Correction (QEC)} & Protects quantum information from noise and decoherence & Reliable communication & Enhancing fidelity in quantum channels (Shor, Steane codes) \\ \hline

\textbf{Quantum Fourier Transform (QFT)} & Quantum analog of DFT, used in signal and phase analysis & Foundational subroutine & Signal processing, spectral graph analysis, pattern recognition \\ \hline

\textbf{Bounded-Error Quantum Polynomial (BQP)} & Decision problems solvable by quantum computers in polynomial time & Complexity class guiding algorithm design & Routing, scheduling, graph optimization \\ \hline

\textbf{Grover’s Algorithm} & Amplitude amplification for searching unsorted data & Quadratic speedup for search problems & Fast pathfinding, network routing, intrusion detection \\ \hline

\textbf{Quantum Walk Algorithms (QWA)} & Quantum analogs of random walks, faster convergence & State space exploration & Routing, clustering, anomaly detection \\ \hline

\textbf{Deutsch-Jozsa Algorithm} & Determines if a function is constant or balanced with one query & Demonstrates quantum speedup & Pattern identification in network traffic \\ \hline

\textbf{QAOA} & Hybrid algorithm alternating between quantum state prep and classical optimization & Approximate solution to combinatorial problems & Traffic flow, load balancing, scheduling in 6G \\ \hline

\textbf{VQE} & Uses variational methods with quantum trial states to find minimum values & Energy or cost function minimization & Resource allocation, optimization tasks in networks \\ \hline

\end{tabular}
\end{table*}

\subsection{Products and Chipsets for Communication and Computing}
\label{subsec:products}


\subsubsection{\glspl{QPU}} The prospects of quantum computing have accelerated efforts to realize fully functional \glspl{QPU} \cite{chi2022programmable}. Recent successes in the development of proof-of-principle \glspl{QPU} have raised the question of how these emerging processors can be integrated into modern high-performance computing systems \cite{britt2017high}. A \ac{QPU} can be defined as the equivalent of a \ac{CPU} in a classical computer. In contrast to a \ac{CPU}, which works with classical bits, a \ac{QPU} uses quantum bits or qubits. As a commercial product, the Forte Enterprise, for example, is proof of IonQ's commitment to pushing the boundaries of quantum computing. \footnote{Online:https://ionq.com/quantum-systems/forte} With its powerful quantum processor and seamless integration into existing data centers, this system offers a versatile and efficient solution for complex computational tasks. Intel has also unveiled a superconducting quantum test chip with enough qubits to potentially enable quantum computing, which is beginning to exceed the practical limits of modern classical computers, as another use case for a data center-based cloud architecture for mobile communications \cite{hsu2018ces}. IBM's Eagle family of processors incorporates more scalable packaging technologies than previous generations. In this architecture, signals pass through multiple chip layers to enable high density \ac{I/O} without sacrificing performance \cite{chow2021ibm}.

\subsubsection{Quantum Time Synchronization} Quantum synchronization offers a promising solution for addressing the challenges faced by modern mobile networks for the use cases such as Autonomous Vehicles,Time-Sensitive Applications, etc \cite{nande2024time}. A source generates a pair of entangled particles, often photons. These particles are inextricably linked, sharing a quantum state regardless of their physical separation. One photon is sent to location A, and the other to location B. At precisely the same moment, both locations measure the state of their respective photons. Due to the entangled nature of the particles, the results of the measurements will always be correlated. This correlation arises from the shared quantum state of the entangled pair. By analyzing the correlation between the measurement results at locations A and B, it is possible to determine the difference in time between the two locations with extraordinary precision \cite{koppenhofer2020quantum}. This is because the correlation is directly tied to the time difference between the measurements. The key advantage of this method lies in the unparalleled accuracy it offers. Quantum entanglement allows for synchronization with sub-picosecond precision, far surpassing traditional timekeeping methods \cite{hong2022demonstration}. For the productwise side, Infleqtion's Tiqker product represents a developing and characterising miniaturised clocks based on both hot and laser cooled atoms. This product can be used to ensure continuous access to precise timing for data centers and telecommunications infrastructures \cite{infleqtion2024}. 


\subsubsection{Quantum Cloud Services} 6G can provide the high-speed, low-latency connectivity needed to support quantum computing applications \cite{wang2022quantum}. IBM Quantum Cloud provides access to IBM's quantum computers and software via a cloud-based platform. It offers a range of quantum computers, including the IBM Quantum System One, which is designed for commercial use. This provides tools for programming quantum computers and simulating quantum circuits through the Qiskit Software Development Kit \cite{majumder2017experimental}. On the other hand, Amazon is introducing Amazon Braket, a cloud service for quantum computers from Amazon Web Services. It provides access to quantum computers from D-Wave, IonQ and Rigetti. It provides tools for programming and executing quantum algorithms. It includes Amazon Braket \ac{SDK}, which provides APIs for programming quantum algorithms. Users can also use existing frameworks and libraries for classical computing \cite{khan2022quantum}. Microsoft Azure Quantum is another cloud service for quantum computing offered by Microsoft. It also provides tools for programming and executing quantum algorithms. Microsoft Azure Quantum offers a flexible and scalable platform for exploring the potential of quantum computing and developing innovative telecommunications applications. It also has the ability to combine quantum computing with classical computing for hybrid applications. This can increase the performance and efficiency of certain tasks \cite{hooyberghs2022azure}. Other notable quantum cloud services include Google Quantum AI, Honeywell Quantum Solutions and Xanadu Quantum Technologies.

Table \ref{tab:characterists_quantum} contains the characteristics of the most important quantum applications in Beyond 5G and 6G networks.

\begin{table*}[htp!]
 \captionsetup{font=sc,  position=above, justification=centering, labelsep=newline,singlelinecheck=true}
\centering
\caption{Characteristics of the most important quantum applications in Beyond 5G and 6G networks}
\label{tab:characterists_quantum}
\scriptsize
\begin{tabular}{|p{2.5cm}|p{3cm}|p{1cm}|p{3cm}|p{3cm}|p{3cm}|}
\hline
\rowcolor{gray!30}
\textbf{Application}                                      & \textbf{Scalability}                        & \textbf{Expected Latency}     & \textbf{Expected Data Capacity}  & \textbf{Other Requirements}                                      & \textbf{Possible Implementations/Examples}                     \\ \hline
\textbf{QKD}       \cite{liu202240,james2023key}              & Hundreds of links and tens of users              & Low                   & Secure key rates up to 1 Mbps  & Dedicated quantum channels, cryogenic systems               & Secure banking, military communication, sensitive data transfers \\ \hline
\textbf{Quantum Sensors for Smart City Infrastructure}   \cite{kamruzzaman2022key,gschwendtner2024quantum}   & Scalable across urban regions        & Medium                   & Low data rates, high precision & Real-time quantum sensor integration with IoT \cite{bhatia2020quantum}      & Quantum-enhanced smart grids, urban monitoring systems    \\ \hline
\textbf{Quantum Computing for Traffic Management} \cite{neukart2017traffic}        & Scalable to city-wide networks         & Low                    & High data processing capacity & Requires integration with edge computing infrastructure     & Real-time traffic monitoring and management               \\ \hline
\textbf{Quantum Encryption for Data Centers} \cite{chen20206g}            & Hundreds of devices                    & Low                    & 100+ Gbps                    & Hybrid quantum-classical encryption protocols                & Securing financial, cloud-based transactions              \\ \hline
\textbf{QML for 6G RAN} \cite{narottama2023quantum,bouchmal2023classical}       & Thousands of devices                   & Medium                   & Data throughput 50 Gbps       & Quantum-enhanced algorithms for resource optimization        & Enhanced resource allocation for 6G networks              \\ \hline
\textbf{Quantum Time Synchronization for 6G Networks} \cite{nande2024time,duong2022quantum}    & Scalable across global networks        & Low                    & Low data rates               & High-precision quantum clocks                                & Timing synchronization in financial markets, defense sectors \\ \hline
\textbf{Quantum Computing for Network Slicing}  \cite{borah2023enhancing}          & Utilize photonic architectures with silicon spin qubits to enable horizontal scalability by adding more modules                       & Medium                   & Capable of achieving high data transfer rates  per slice             & High computational power, QPU integration                    & Dynamic slicing for enhanced mobile broadband             \\ \hline
\textbf{Quantum Cloud Computing for Edge Applications} \cite{mastroianni2023quantum}    & Distributed architecture to enable horizontal scalability across multiple quantum modules                    & Medium                   & Ability to handle approximately 2.5 exabytes of data generated daily \cite{einfochips2024}  & Integration with quantum cloud platforms, QPU-powered edge   & AI-powered cloud applications for edge devices and reduce workloads in data centers by up to 12.5\%, enhancing efficiency in handling large datasets \cite{quantuminsider2024}            \\ \hline
\textbf{Quantum-Enhanced 6G Core Network Operations} \cite{rozenman2023quantum}      & High potential for handling massive data loads and complex computations.   & Medium                   & High-speed data processing    & Quantum computing cores for fast network decision-making     & Optimization of network traffic, dynamic resource allocation \\ \hline
\textbf{Quantum-Enhanced Holographic Communication} \cite{zhou2023towards}       & Higher scalability with a wider range of applications      & Medium                   & Higher or lower depending on technological advancements and specific implementations & Real-time processing via quantum-assisted AI                 & Real-time holographic communications for remote collaboration \\ \hline
\textbf{Quantum Machine Learning for Cybersecurity} \cite{said2023quantum}       & Scalable across national security systems & Low                  & High-speed data analysis      & Quantum-enhanced cryptographic frameworks                    & Threat detection, anomaly identification in critical infrastructures \\ \hline
\textbf{Quantum-Based uRLLC} \cite{zaman2023quantum} & Scale to a significant degree of connected devices per base station & Low              & Orders of magnitude higher data rates                 & Quantum error correction, high-bandwidth connections          & Autonomous driving, smart grid management                  \\ \hline
\end{tabular}
\vspace{-5mm}
\end{table*}

\section{Quantum technologies for 6G RAN Technologies}
\label{sec:quantum}




\textcolor{black}{This section builds directly on the foundational concepts discussed in Section \ref{sec:background}, particularly the principles of quantum communication (Section \ref{subsec:quantumtech}), hybrid quantum-classical systems (Section \ref{subsec:hybrid}), and hardware advancements (Section \ref{subsec:products}). These foundations are essential for understanding the role of quantum technologies in the evolution of 6G \ac{RAN}. Quantum \ac{MIMO} architectures, for example, rely on entanglement and quantum-based signal processing to increase transmission security and spectral efficiency. Similarly, quantum-assisted reflective surface technologies and cooperative beamforming approaches use quantum principles to enable wireless communication with low latency, security and high capacity — critical attributes for future RAN deployments in ultra- dense and heterogeneous environments.}

\subsection{Introduction to Quantum Technology in 6G RAN Communication}

The shift from 5G to 6G wireless networks marks the beginning of a new chapter in communication, marked by significant advances and ambitious goals. This evolution encompasses a diverse range of themes and technologies that contribute to the expansive landscape of 6G networks. The vision for 6G extends beyond the capabilities of 5G, with a focus on achieving multi-terabyte per second data rates \cite{zhang20196g}. The evolving 6G framework combines space, air, terrestrial, and underwater networks, with the goal of providing seamless and boundless wireless connectivity everywhere \cite{zhang20196g}. Central to this transformation are AI and machine learning, which are instrumental in realizing autonomous networks in the 6G era \cite{zhang20196g} \cite{chowdhury20206g}. These technologies enable advancements such as supermassive Multiple-Input, Multiple-Output \ac{MIMO} arrays, terahertz communications, orbital angular momentum multiplexing, holographic beamforming, large intelligent surfaces, laser and visible-light communications, quantum communications and computing, blockchain-based spectrum sharing, molecular communications, and the Internet of Nano-Things \cite{zhang20196g}.

6G addresses the limitations of 5G by offering enhanced system capacity, higher data rates, reduced latency, improved security, and superior quality of service \cite{chowdhury20206g} \cite{schwenteck20236g}. It leverages cutting-edge technologies, including AI, terahertz communications, quantum communications, optical wireless technology, and blockchain \cite{chowdhury20206g}. The demand for 6G is driven by the need for private networks in machine-to-machine communication, which require cost-effective and user-friendly configurations \cite{schwenteck20236g}. Additionally, 6G aims to tackle modern challenges such as those posed by the Metaverse, with a focus on ultra-low latency, data handling, energy efficiency, and security \cite{schwenteck20236g}. Technologies like Post-Shannon, Molecular, and Quantum communication are expected to revolutionize human-to-machine interactions, making them more intuitive and efficient \cite{schwenteck20236g}.

\begin{figure}[htp!]
  \centering
  \includegraphics[width=0.5\textwidth]{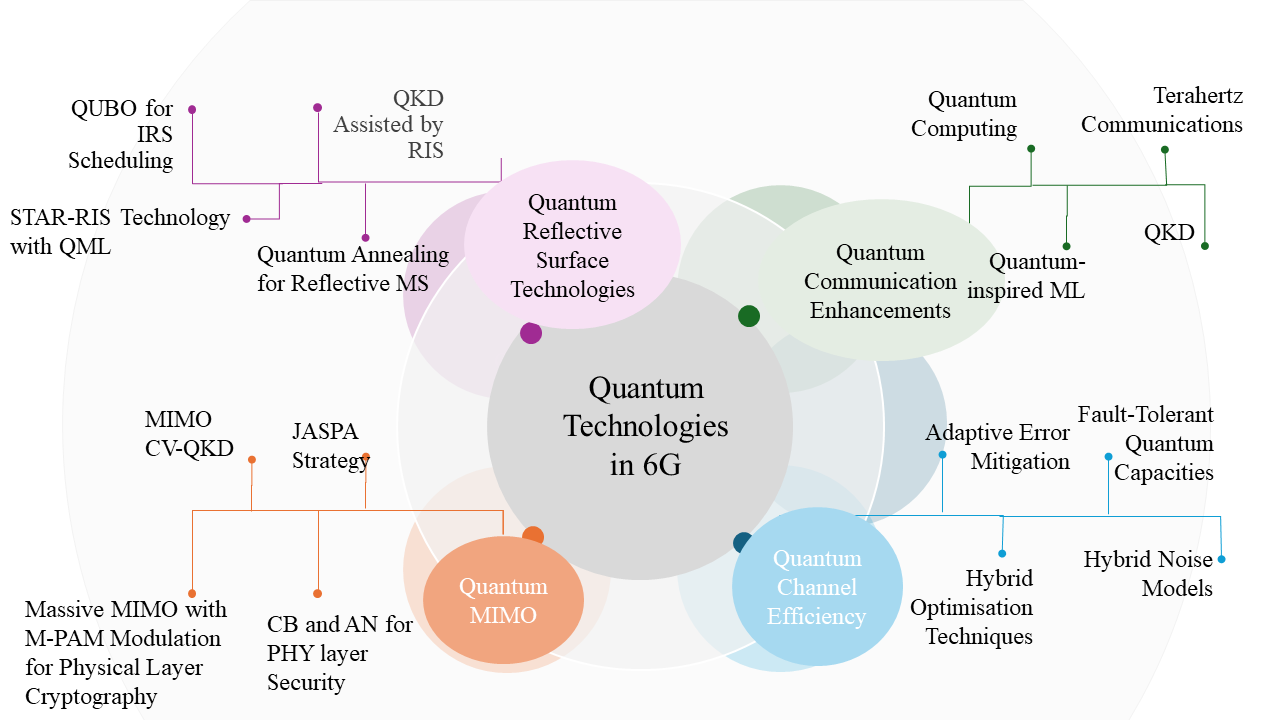}
  \caption{Quantum technologies integrated  \ac{RAN} 6G Technology.
  }
    \label{RAN_fig}
    \vspace{-.5cm}
\end{figure}

Beyond its communication capabilities, 6G also emphasizes advancements in security technologies \cite{chen20206g}. Artificial Intelligence, Quantum Computing, Communications, and Blockchain are anticipated to enhance the security of 6G networks \cite{chen20206g}. These technologies aim to provide broader coverage and an improved user experience, marking the beginning of a secure and advanced 6G era \cite{chen20206g}. Quantum technology, in particular, plays a crucial role in 6G wireless communication \cite{akyildiz20206g}. Quantum communications, grounded in quantum mechanics, promise tamper-proof data transfer and the simultaneous encoding and transmission of multiple data streams, making it a key enabling technology \cite{akyildiz20206g}. Nonetheless, challenges such as \ac{QEC} and entanglement distribution needs more research to completely realizing the potential of quantum technologies \cite{akyildiz20206g}. 6G not only emphasizes enhanced networking performance but also highlights the importance of secure, intelligent network architecture and a human-centric approach \cite{bandi2022review}. It underscores the need for high-performance networking metrics, secure network infrastructure, and a focus on user experience, driving the development of 6G wireless networks \cite{bandi2022review}. Governments and industry leaders have already embarked on 6G projects, ensuring a vibrant and promising future for wireless communication \cite{bandi2022review}. In the following sections, Quantum  \ac{RAN} technologies have been discussed with their implementation and discussion and Fig. \ref{RAN_fig} highlights some of the research technologies.  In this figure, we used the following terms: Quantum Unconstrained Binary Optimization (QUBO), \ac{QML}, Continuous Variable Quantum Key Distribution (CV-QKD), \ac{QKD}, Joint Antenna Selection and Power Allocation (JASPA), \ac{RIS}, and Metasurfaces (MS).

\subsection{Quantum Technologies \textcolor{black}{for} 6G}


The shift from 5G to 6G networks emphasizes key themes such as sustainability, equity, security, and the exponential growth of wireless data demands \cite{khanh2023innovative}. Central to this evolution is the integration of quantum technologies alongside AI, terahertz communication, visible light communication, and blockchain, which collectively promise higher data rates, lower latency, and improved network reliability for applications like smart cities, autonomous vehicles, and virtual reality  \cite{vu2023potential}. Quantum communication stands out as a critical technology in 6G, enhancing security through post-quantum cryptography and quantum key distribution. These innovations aim to address the vulnerabilities exposed in 5G networks and ensure secure data transfer in 6G \cite{uysal2022data}. Quantum technologies are also being explored in healthcare services and play a pivotal role in optimizing network performance through quantum-inspired machine learning algorithms and real-time resource allocation \cite{cornet2022overview} \cite{duong2022quantum}. In addition to security, 6G aims to achieve global coverage through integrated satellite-terrestrial networks, leveraging machine learning and quantum computing for optimized multibeam designs and real-time support for critical services \cite{duong2023machine}. Quantum communication, along with AI-driven optimization, is expected to revolutionize connectivity, moving beyond human-to-machine interactions toward intelligent system integration \cite{bandi2022towards}. Conclusively, the shift to 6G underscores the pivotal impact of quantum technologies, especially in enhancing security and optimising networks, both of which are crucial for achieving the ambitious performance targets of future wireless communication systems.

\subsection{Quantum MIMO}

Quantum \ac{MIMO} combines the security benefits of quantum communication with the efficiency of \ac{MIMO} technology, addressing the weaknesses of traditional cryptography in the age of quantum computing. By using techniques such as \ac{QKD} and advanced MIMO configurations, it promises increased security, improved secret key rates and greater transmission distances. The main challenges and potential solutions as well as implementation details are discussed in the following sections.

\subsubsection{Implementations} 

\textcolor{black} {The concept of physical-layer cryptography, particularly leveraging massive MIMO systems for secure communication, is explored in \cite{dean2017physical}. Traditional cryptography is vulnerable to quantum computing threats, but utilising massive MIMO with M-PAM modulation makes decoding as complex as solving lattice problems, which are computationally difficult for both classical and quantum computers. Recent advancements in quantum communication, especially with \ac{THz} and millimeter waves (mm-waves), have shown promise for enhancing the security of QKD systems. Studies on MIMO Continuous Variable QKD (CV-QKD) techniques explore frequency ranges from 100 GHz to 1 \ac{THz}. One study \cite{kundu2023mimo} introduces a MIMO CV-QKD system within a "restricted" eavesdropping scenario, where the eavesdropper (Eve) can only intercept a fraction of the lost photons. This model offers a more optimistic Secret Key Rate (SKR) and improved security for real-world quantum communication.}

\textcolor{black} {Another study \cite{zhang2023millimeter} demonstrates the feasibility of mm-wave and THz QKD over long distances using MIMO configurations. Operating at cryogenic temperatures, such as 4 K, and using a 1024 × 1024 MIMO antenna setup, secure communication over distances exceeding 5 km at 100 GHz was achieved, showing the potential of this technology for long-range secure systems. In the realm of 6G communications, \cite{jang2023cooperative} tackles the security challenges present in extensively connected IoT networks. The paper introduces a Cooperative Beamforming (CB)-based Physical Layer Security (PLS) scheme enhanced with Artificial Noise (AN) to protect against eavesdropping. The study also provides a closed-form secrecy rate expression, considering Channel State Information (CSI) errors, and demonstrates the scheme's effectiveness in improving secrecy rates. To enhance the utilisation of time-frequency resources and physical layer security in full-duplex massive MIMO networks, \cite{gao2021joint} suggests a Joint Antenna Selection and Power Allocation (JASPA) strategy. By treating simultaneous transmissions as same-frequency interference, interception is complicated. The study introduces a \ac{QBSA} to optimize the JASPA setup, achieving superior secrecy performance compared to existing methods.}

\subsubsection{Discussion}

Recent studies on QKD at millimeter-wave (mm-wave) and THz frequencies have highlighted both the potential and challenges of these technologies in secure communication. Research has shown that using MIMO technology significantly enhances secure transmission distances compared to Single-Input Single-Output (SISO) techniques. An increase in the number of antennas in a MIMO setup directly correlates with extended transmission distances and improved SKR, demonstrating the scalability of these systems for real-world applications \cite{zhang2023millimeter}. However, achieving this improvement presents substantial technical challenges. For instance, operating at cryogenic temperatures below 50 K is necessary to maximize the efficiency of solid-state components like sources and detectors, which are crucial for achieving high SKRs and extending communication distances. The need for advanced cooling infrastructure creates a considerable barrier to widespread adoption, highlighting the importance of future advancements in cryogenic technology for the broader implementation of mm-wave and THz QKD systems. These technologies have potential applications in areas such as intersatellite communication, deep space networks, and secure indoor communications \cite{zhang2023millimeter}.

Additional insights into QKD systems under realistic eavesdropping conditions indicate that when the eavesdropper, Eve, has limited access to the communication channel, information leakage—as measured by the Holevo information—is diminished, thereby improving the secret key rate (SKR). Coherent-state-based CV-QKD protocols are especially effective in these restricted eavesdropping settings. These protocols are simpler to implement and more robust in practical situations where the connection between Alice and Eve is lossy. While increased squeezing may boost SKR under unrestricted eavesdropping, it leads to higher information leakage and a reduced SKR in restricted scenarios, making coherent-state-based protocols more suitable for practical applications \cite{kundu2023mimo}. In addition to QKD improvements, using same-frequency interference generated by simultaneous transmissions at both the base station and user ends has emerged as an effective strategy for enhancing physical layer security. This method complicates the task for eavesdroppers by creating an interference-rich environment, providing a novel security application for wireless networks \cite{gao2021joint}. Additionally, the JASPA strategy has been shown to lower energy consumption and enhance resource utilisation in Co-time Co-frequency Full-Duplex (CCFD) massive MIMO networks. This effective combination of antenna selection and power allocation is vital for optimising both secrecy performance and the system’s quality of service (QoS) \cite{gao2021joint}. Furthermore, the Quantum-Based Swarm Algorithm (QBSA) has outperformed other intelligent algorithms in optimising the JASPA strategy, highlighting the significant impact of advanced algorithms on improving the security and performance of wireless communication systems. It is recommended to extend these methods to ultra-dense networks, integrating them with ultra-massive MIMO and other emerging technologies to further bolster the security and efficiency of wireless communications \cite{gao2021joint}.

\subsection{Quantum-Enhanced Communication Systems and Reflective Surface Technologies}

Recent advancements in quantum communication systems have significantly improved efficiency and security through the integration of quantum-based optimisation techniques and reconfigurable intelligent surfaces (RISs). This research explores innovative methods for managing phase configurations and channel estimation, utilising quantum annealing for reflective metasurfaces and sophisticated quantum machine learning frameworks. These developments address key challenges in phase modulation and secure quantum key distribution, setting the stage for enhanced communication networks in next-generation wireless technologies.
\subsubsection{Implementations}
\textcolor{black} {A variety of innovative approaches have advanced quantum communication systems, with RIS and quantum-based optimisation playing central roles. \ac{QA} has been applied to optimise reflective metasurfaces (MS) through the use of spin-chain Hamiltonians \cite{9665281}. By employing the D-Wave 2048-qubit quantum optimiser, the phase modulation of binary and quadriphase MS is enhanced, showing potential for RIS optimisation in forthcoming 5G/6G networks. In the realm of quantum computing hardware, flicker noise (1/f noise) for 45-nm RFSOI NFETs is characterised at cryogenic temperatures \cite{10433783}. This analysis, essential for quantum circuits, highlights how carrier number fluctuation drives noise behaviour, which remains stable even at low temperatures. Optimisation techniques extend into 6th-generation (6G) cellular systems through IRS allocation scheduling \cite{9555241}. Quantum Unconstrained Binary Optimisation (QUBO) is employed to improve energy efficiency and communication scheduling in multi-user environments, outperforming traditional methods like OFDMA. \ac{FSO} with the assistance of Intelligent Reflecting Surfaces (IRS) have been suggested to improve QKD \cite{10289124}. This system accommodates both discrete variable QKD (DV-QKD) and continuous variable QKD (CV-QKD), facilitating quantum-secure key exchange over a distance of 2 km, even in non-line-of-sight situations. STAR-RIS (Simultaneously Transmitting and Reflecting RIS) technology further enhances RIS capabilities by enabling both signal reflection and transmission \cite{10293918}. A modular quantum machine learning (QML) framework is used to improve channel estimation, reducing complexity and communication overhead. Finally, an RIS-aided multi-user QKD system was introduced to mitigate performance degradation caused by obstacles in FSO environments \cite{10168920}. The system optimises the RIS phase matrix, securing communications even when trusted nodes are compromised by eavesdropping. These innovations in quantum optimisation and intelligent surfaces collectively elevate the security, efficiency, and performance of quantum communication systems across diverse applications.}

\subsubsection{Discussion}

The collective findings across the studies present significant advancements in the field of quantum communication, specifically through optimisation of MSs and RISs. The utilisation of the Ising Hamiltonian model for MS optimisation introduces a new paradigm in Electromagnetic (EM) optimisation tasks. By shifting away from traditional artificial intelligence approaches, this technique leverages QA to optimise scattered wave power, demonstrating its potential for next-generation wireless communication systems \cite{9665281}. Future research is anticipated to investigate multi-phase RISs, the joint optimisation of reflection amplitude and phase, as well as the effects of mutual coupling. Additionally, the examination of flicker noise in 45-nm RFSOI NFETs offers valuable insights for applications in quantum computing. The behaviour of normalised noise remains temperature-independent, with a U-shaped frequency exponent response at cryogenic temperatures \cite{10433783}. It highlights the role of non-uniform energy trap distributions at low temperatures.
This has implications for improving Si qubits and CMOS devices in quantum architectures.  In the context of 6G systems, the proposed IRS allocation scheduling method, solved via Quantum Unconstrained Binary Optimisation (QUBO), offers substantial improvements in communication efficiency by reducing time slot usage and increasing IRS array gain \cite{9555241}. The potential of quantum computing to handle combinatorial optimisation for larger networks as quantum hardware evolves is promising, although challenges such as user mobility and overheads in pilot signal transmission remain unresolved and warrant further exploration. \ac{QKD} systems benefit significantly from IRS assistance, particularly in \ac{FSO} communication. DV-QKD outperforms CV-QKD at longer transmission distances, while CV-QKD facilitates with better SKR over shorter distances, suggesting that the protocol selection depends on particular communication requirements \cite{10289124}. However, the reliance on perfect channel knowledge for phase optimisation presents a challenge, and further work is needed to incorporate channel estimation errors into SKR analysis. The modular \ac{QML} framework introduced in STAR-RIS systems shows promising results in improving channel estimation by segmenting tasks into denoising and extrapolation modules \cite{10293918}. This innovation reduces noise and facilitates accurate channel prediction, marking a step forward in the optimisation of RIS environments.

Finally, the integration of a trusted RIS in \ac{QKD} networks enhances security by maintaining encryption consistency across multiple quantum sublinks, even in the event of compromised trusted nodes (TNs) \cite{10168920}. This approach is particularly effective in terrestrial environments where obstacles disrupt line-of-sight communication, ensuring that the overall security of the system remains robust despite external vulnerabilities. These advancements in quantum-based optimisation techniques, intelligent surfaces, and secure communication protocols offer critical developments for the future of quantum communication systems, although challenges related to channel estimation, user mobility, and hardware scalability remain areas for further research and refinement.

\subsection{Quantum Channel Efficiency}
This section explores various advancements aimed at enhancing the efficiency and reliability of quantum communication channels. The research reviewed covers diverse areas, including fault-tolerant quantum communication, noise management, adaptive error mitigation, and hybrid optimisation techniques. These studies tackle essential challenges including channel noise, fault tolerance, and resource limitations, offering valuable insights for the practical deployment of quantum communication systems. By examining how these techniques work together, we gain a clearer understanding of the strategies being developed to optimise quantum channels for more robust and scalable communication.
\subsubsection{Implementation}
\textcolor{black} {In quantum communication, ensuring reliable data transmission in the presence of noisy channels and other uncertainties is a critical challenge addressed through various innovative strategies across several research works. The work in \cite{9761242} addresses the issue of noise in quantum gates and channels by developing fault-tolerant quantum capacities. By merging fault-tolerant quantum computation techniques with quantum Shannon theory, this study introduces capacities that allow near-optimal communication rates with minimal errors, even when circuits are prone to gate errors. This fault-tolerant capacity, dependent on gate error probability, establishes a threshold beyond which reliable quantum communication can be sustained, effectively demonstrating error resilience in both classical and quantum channels. This thresholding approach is crucial for achieving error-free communication within a specific error rate tolerance, which links closely to efforts in other studies to manage noise more effectively.}

\textcolor{black} {Building on the challenge of noise management, \cite{10403910} explores hybrid quantum noise environments, combining quantum and classical noise models to derive channel capacities. This study provides a mathematical framework for calculating quantum channel capacity by modelling both classical Additive White Gaussian Noise (AWGN) and Poisson-distributed quantum noise. The application of Gaussian Mixture Models (GMMs) to characterise noise allows for a statistical representation of quantum channel behaviour, resulting in the conclusion that Gaussian signalling is more effective for optimising data rates in these noisy settings. This complements the findings in \cite{9761242} by confirming that, even with noisy quantum environments, optimal capacity can be achieved through precise modelling and signal optimisation. In parallel, \cite{10645687} addresses the challenge of nonstationary noise, which varies over time, complicating reliable quantum computations. The authors implement Probabilistic Error Cancellation (PEC) using Bayesian inference to dynamically adjust noise estimates, significantly improving both accuracy and stability in quantum systems. This approach complements the fault-tolerant strategies seen in \cite{9761242} by tackling temporal noise fluctuations, whereas \cite{9761242} focuses on error thresholds. Furthermore, the scalability challenges raised in \cite{10645687}, such as the high resource demands of error mitigation circuits, echo the resource management strategies seen in later research. }
\textcolor{black} {To tackle the aforementioned resource challenges, the research in \cite{10549913} introduces sophisticated hybrid optimisation techniques for systems based on Quantum-dot Cellular Automata (QCA). By employing Quantum-inspired Genetic Algorithms (QIGA) alongside Reinforcement Learning (AGA-RL), the system can dynamically adapt to varying channel conditions, thereby ensuring both fault tolerance and efficient resource utilisation. Techniques like Error-Correcting Codes (ECC) and Dynamic Resource Allocation (DRA) optimise the balance between resource usage and communication quality, aligning with the modular noise mitigation strategy proposed in \cite{10645687}, where scalable models reduce the overhead of error mitigation. Thus, these studies collectively contribute to optimising quantum communication systems by addressing noise management from different angles: fault-tolerant capacities \cite{9761242}, hybrid noise environments \cite{10403910}, adaptive error correction \cite{10645687}, and hybrid optimisation for fault tolerance and resource management \cite{10549913}. Together, they establish a comprehensive framework for reliable, scalable, and efficient quantum communication under various noisy conditions.}
\subsubsection{Discussion}
The studies collectively advance quantum communication systems by tackling key challenges such as noise management, error resilience, and resource optimisation. The research in \cite{9761242} achieves near-ideal quantum communication rates even in noisy environments, demonstrating practical fault tolerance with minimal overhead. This breakthrough raises questions about the universality of these results and suggests potential for further exploration in private and multi-party quantum communication. The work in \cite{10403910} develops a robust statistical framework for hybrid quantum noise environments, revealing that Gaussian inputs are optimal for maximising data rates. This finding is significant but does not extend to more complex quantum noise models, indicating a need for further research. In \cite{10645687}, Bayesian inference improves the stability and accuracy of probabilistic error cancellation, showing substantial gains. However, scalability issues due to increased resource costs are noted, with suggestions for modular noise modelling to enhance efficiency. Finally, \cite{10549913} introduces hybrid optimisation techniques for Quantum-dot Cellular Automata (QCA)-based systems, including quantum-inspired algorithms and dynamic resource allocation. These methods improve adaptability, error correction, and resource efficiency, with deep reinforcement learning enhancing convergence speed and solution quality.

Overall, these studies offer complementary solutions to the fundamental challenges in quantum communication, providing a strong basis for future developments. A summarised version of Section \ref{sec:quantum} is given in Table \ref{tab:RAN}.

\begin{table*}[htp!]
\captionsetup{font=sc}
\centering
\caption{Quantum Technologies in 6G RAN.}
\scriptsize
\label{tab:RAN}
\begin{tabular}{|p{3cm}|p{6cm}|p{6cm}|p{1.2cm}|}
\hline
\rowcolor{gray!30}
\textbf{Quantum Technology} & \textbf{Description} & \textbf{Application in 6G RAN}                 & \textbf{Reference} \\ \hline
\textbf{Quantum Annealing for RIS}                & Optimisation of RIS phase modulation using quantum annealing techniques, improving electromagnetic wave manipulation.           & Enhances the signal reflection and transmission efficiency in 6G RANs, boosting energy efficiency and communication capacity.   &  \cite{9665281}  \\ \hline
\textbf{Flicker Noise Characterisation in Quantum Hardware}     & Analysis of flicker noise (1/f noise) in quantum circuits at cryogenic temperatures for stable operations.    & Improves quantum hardware reliability in 6G RAN base stations, especially for quantum computing and optimisation tasks.  &  \cite{10433783}        \\ \hline
\textbf{QUBO}          & A quantum optimisation method that improves resource scheduling for RIS allocation in multi-user scenarios.            & Optimises power usage, channel allocation, and communication scheduling for massive MIMO and RIS-assisted 6G RANs.  & \cite{9555241}  \\ \hline
\textbf{IRS-Assisted FSO Quantum Communication}& Enhances QKD protocols by employing IRS in FSO systems to secure key exchanges in non-line-of-sight conditions.  & Provides secure and long-distance communication links between base stations and users through QKD, ensuring robust quantum security in RAN.  & \cite{10289124}    \\ \hline
\textbf{STAR-RIS with QML}&Quantum ML models are used to enhance RIS capabilities for simultaneously transmitting and reflecting signals.&Reduces channel estimation complexity and improves communication overhead in 6G RAN, ensuring reliable signal transmission and reflection.&\cite{10293918}\\ \hline
\textbf{Trusted RIS-Aided Multi-User QKD}&A trusted RIS system that secures communication in multi-user quantum networks, even in compromised environments.&Enables multi-user QKD for secure data transmission in dense urban environments, a key feature of 6G RAN deployments.&\cite{10168920}\\ \hline
\textbf{Fault-Tolerant Quantum Communication}&Develops quantum communication systems with high error resilience, even in noisy environments, using fault-tolerant quantum capacities.&Ensures reliable, error-free data transmission between RAN elements, increasing the robustness of 6G quantum communication networks.&\cite{9761242}\\ \hline
\textbf{Hybrid Noise Management for Quantum Channels}&Combines classical and quantum noise models to derive optimal channel capacities, improving data transmission in noisy channels.&Optimises quantum channel performance in 6G RAN, ensuring higher data rates and more reliable communication in noisy urban environments.&\cite{10403910}
\\ \hline
\end{tabular}
\vspace{-0.5cm}
\end{table*}

\section{Quantum technologies for 6G Core,  Edge and Transport}
\label{sec:core}



\begin{figure}[htp!]
    \centering
    \includegraphics[width=0.8\linewidth]{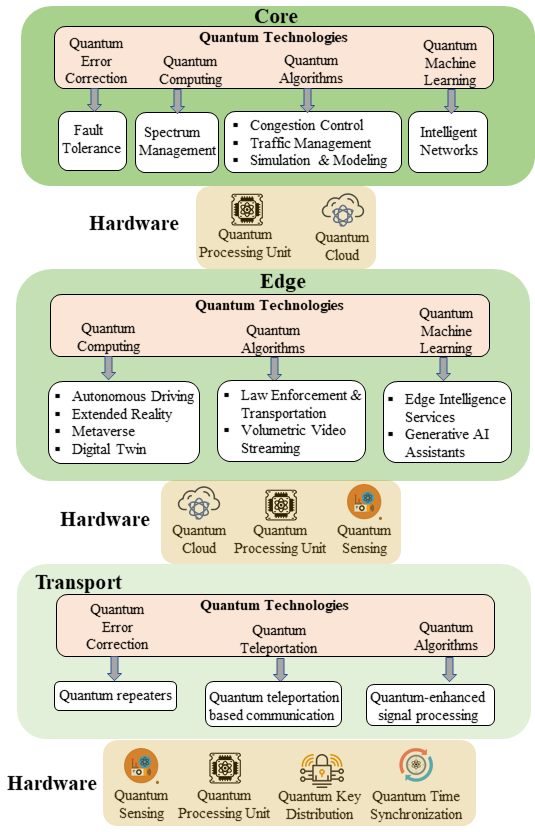}
    \caption{Quantum technologies and supported applications for the realization of the state-of-the-art implementations for 6G.} 
    \label{fig:QC_tech}
    \vspace{-.7cm}
\end{figure}

\textcolor{black}{The core, edge, and transport layers of 6G networks present unique challenges that quantum computing can address. This section extends the discussion from Section \ref{sec:background}—particularly hybrid quantum-classical computing (Section \ref{subsec:hybrid}), quantum optimization for network control (Section \ref{subsec:QC}), and hardware developments (Section \ref{subsec:products}). Quantum algorithms for routing, resource allocation, and traffic optimization, supported by scalable hardware solutions, are vital for improving computational performance and decision-making at the network core. At the edge, decentralized quantum processing enables real-time analytics and low-latency response for applications such as autonomous vehicles and industrial \ac{IoT}. Furthermore, quantum-assisted transport protocols and quantum repeaters (as introduced in Section \ref{subsec:QC}) play a critical role in supporting ultra-reliable, secure, and long-range data transmission across distributed 6G infrastructures.} As 6G technologies and enablers have only recently been an ongoing project and a call for proposals \cite{6G-NIST}, some technologies have been suggested for incorporation into the 6G Core and Edge. For example, quantum technologies can be used to facilitate some key functions within the core and edge components of the 6G network. Fig. \ref{fig:QC_tech} presents the mapping of quantum technologies and current quantum hardware on the related quantum applications.

\subsection{Quantum Computing at the 6G Core}
\textcolor{black}{The 6G core is expected to leverage cloud-native principles such as containerization, microservices, and service-based architecture to ensure scalability, flexibility, and resilience. These components are dynamically orchestrated using AI-driven platforms that optimize service placement, resource allocation, and policy enforcement. A revolutionary aspect is the anticipated integration of quantum-aware and high-performance computing resources. With the rise of quantum computing, the 6G core may utilize quantum co-processors for complex tasks such as network optimization, cryptographic key distribution, and traffic forecasting. Quantum-assisted algorithms can work alongside AI models to solve computational bottlenecks in real time, especially in scenarios involving massive device connectivity and dynamic spectrum management. The 6G core can become a computational fabric—one with a quantum-enhanced, and deeply embedded within the network infrastructure. It can support proactive network behavior, self-healing mechanisms, and service-level guarantees that adapt in real time to user mobility, demand fluctuations, and environmental contexts.}

\textit{Traffic Management}: \textcolor{black}{A primary application of quantum computing in this domain lies in solving complex optimization problems that underpin resource allocation and routing decisions. Algorithms such as the \ac{QAOA} and quantum annealing are particularly suited for addressing NP-hard challenges that involve multi-variable constraints and large decision spaces. In a dynamic network scenario, these algorithms can be applied to optimize bandwidth distribution, buffer management, and routing across multiple slices, considering real-time traffic density, user mobility, and QoS requirements \cite{li2020quantum}. The \ac{QAOA} is a gate-based, variational quantum algorithm specifically designed to solve combinatorial optimization problems, many of which fall under the NP-hard class. It works by mapping the problem into a cost Hamiltonian, where the lowest energy state corresponds to the optimal or near-optimal solution. A second mixing Hamiltonian is used to explore the space of possible configurations. These two Hamiltonians are alternated in a parameterized quantum circuit, and the circuit parameters are optimized using a classical outer loop to minimize the expectation value of the cost function. Thus, QAOA is particularly effective for problems that include a broad range of network scenarios like optimal link selection, interference minimization, or congestion-aware traffic scheduling. As the depth of the circuit increases, QAOA generally yields better approximations of the optimal solution. In parallel, quantum annealing, employed by systems like D-Wave, offers another approach to solving NP-hard problems. Rather than using discrete gate operations, quantum annealing continuously evolves a quantum system from a known, simple initial ground state toward the ground state of a complex cost Hamiltonian that encodes the optimization problem. This process allows the system to explore multiple minima simultaneously and escape local optima more effectively than classical simulated annealing. In network contexts, quantum annealing is useful for solving large-scale routing, channel assignment, and load balancing problems. For example, selecting the most efficient path in a highly congested multi-node wireless mesh network can be modeled and minimized using a quantum annealer, even in real-time or near-real-time applications. }

\textcolor{black}{Together, QAOA and quantum annealing present promising solutions to NP-hard problems in 6G networks, offering faster convergence, better approximations, and scalability. As quantum hardware matures, these methods are expected to be integrated into orchestration systems for proactive and predictive control of complex networking environments \cite{bhatia2020quantum}. Their ability to compute high-quality solutions for large-scale, dynamic optimization problems may become essential for ensuring QoS, minimizing congestion, and optimizing end-to-end latency in future intelligent communication systems.}


\textit{Congestion Control}: \textcolor{black}{Proactive congestion control in 6G networks is significantly advanced by the incorporation of quantum-enhanced reinforcement learning (Q-RL) within network digital twins. These digital twins are real-time, high-fidelity virtual replicas of the physical network infrastructure, continuously synchronized with live network telemetry, user behavior, mobility patterns, and service demands. Their primary function is to simulate "what-if" scenarios under diverse traffic conditions, enabling predictive insights and policy testing without disrupting real-world operations \cite{qu2022temporal}. By embedding Q-RL agents in these virtual environments, 6G networks gain the ability to learn, evaluate, and refine adaptive congestion control strategies in a continuous feedback loop. Quantum-enhanced reinforcement learning merges classical Q-learning or Deep Q-Networks with quantum computational primitives—such as quantum state superposition and entanglement—to accelerate the exploration and policy convergence in large and complex state-action spaces. In the context of 6G, where a single node may face hundreds of potential actions (e.g., rerouting, spectrum switching, buffer resizing), Q-RL allows the agent to consider multiple paths simultaneously and evaluate expected rewards faster than classical agents. For instance, when congestion is predicted in a certain cell or slice, the quantum agent can proactively learn and execute actions such as rerouting traffic across alternative low-latency paths, dynamically reconfiguring spectrum allocations (e.g., switching from sub-6 GHz to mmWave or THz bands), or gracefully degrading non-critical flows based on pre-learned QoS hierarchies. Moreover, quantum-enhanced agents excel in non-stationary environments, which is a hallmark of 6G scenarios involving ultra-mobile users (e.g., UAVs, V2X), network slicing dynamics, and rapidly shifting service topologies. Q-RL can incorporate probabilistic transition models and maintain robust performance even when network conditions deviate from learned patterns, something classical reinforcement learning struggles with.}


\textit{Spectrum Management}: \textcolor{black}{One of the most direct applications of QC in spectrum management is in spectrum allocation and sharing, which are NP-hard optimization problems. Allocating spectrum bands to a large number of users or devices (e.g., in cognitive radio networks, V2X, or IoT) while avoiding interference and maximizing spectral efficiency often involves solving constraint satisfaction problems or graph coloring formulations. QAOA and quantum annealing are particularly suited for solving such combinatorial problems by finding high-quality solutions in large, complex search spaces \cite{liu2016optimization}. For example, QAOA can model spectrum assignment as a graph where nodes represent users and edges represent interference relationships, and it can optimize color (frequency) assignments to minimize conflicts \cite{zhang2022quantum}. \ac{QML} models, such as quantum neural networks or variational quantum classifiers, can be trained to detect occupancy patterns across time and space, identify idle channels, and predict spectrum availability with greater accuracy than classical models. This is particularly valuable in dynamic spectrum access and spectrum-as-a-service scenarios, where real-time decision-making is critical. These models can process massive volumes of telemetry data (e.g., signal strength, location, usage history) faster due to quantum-enhanced feature space representation, enabling cognitive radio systems to learn and adapt quickly to changing spectral environments. Another dimension is the potential use of quantum-enhanced reinforcement learning in self-organizing RANs. Here, quantum RL agents can learn optimal spectrum access policies based on feedback from the environment, adapting to congestion, interference, and regulatory rules in real time.  These agents may operate at the base station or edge controller level to coordinate spectrum usage among users and services, especially in unlicensed or shared bands. Lastly, as terahertz and optical wireless communication bands are introduced in 6G, the complexity of spectrum management increases further due to atmospheric absorption, beam alignment, and hardware constraints. Quantum algorithms can help manage this complexity by enabling multi-objective optimization—balancing bandwidth efficiency, energy consumption, and interference minimization simultaneously \cite{alsaedi2023spectrum}.}

\textit{Fault Tolerance}: It is crucial to guarantee the dependability and durability of 6G networks, particularly given their growing significance for many facets of the business and society. Traditional fault tolerance mechanisms, while effective to a certain extent, often rely on redundancy and pre-defined recovery procedures. \ac{QC} offers a novel approach to enhance fault tolerance by leveraging \ac{QEC} techniques. The purpose of \ac{QEC} is to safeguard quantum data against mistakes resulting from noise and decoherence, which are intrinsic difficulties in quantum systems. By encoding information across multiple qubits and utilizing sophisticated error detection and correction algorithms, Quantum computations can be made much more reliable and fault-tolerant by using \ac{QEC} \cite{liu2023enabling}. The network can detect and repair mistakes in real-time, assuring ongoing operation even in the face of unforeseen disruptions, by encoding key data and control signals using \ac{QEC}. Moreover, QC could enable the development of self-healing networks that can autonomously identify and recover from faults. By continuously monitoring network health and performance metrics, QC algorithms can detect anomalies and potential failures before they cause significant disruptions. This proactive approach could lead to faster fault recovery, minimized downtime, and improved overall 6G network reliability. Therefore, the integration of \ac{QEC} techniques into communication networks could significantly enhance their fault tolerance and resilience \cite{egan2021fault}. In the event of unplanned disruptions, QC could assist guarantee that networks continue to function and provide uninterrupted services by safeguarding vital components and enabling proactive fault recovery. This enhanced reliability is essential for supporting critical 6G infrastructure and enabling mission-critical applications.

\textit{Simulation and Modeling:} The design and optimization of complex communication networks, especially those incorporating cutting-edge technologies like 6G, often rely on accurate simulations and models to predict performance, identify potential bottlenecks, and evaluate different design choices. Classical simulation methods, while valuable, can be computationally intensive and may struggle to capture the full complexity of 6G networks, particularly when considering the intricate interplay of various network elements and traffic patterns \cite{zhou2020limits}. \ac{QC} offers a transformative approach to network simulation and modeling. QC simulators are able to effectively simulate complicated network scenarios and make impressive predictions about their behavior of the network by utilizing the capabilities of quantum computing. Due to quantum computation's intrinsic parallelism, a large number of potential network states and traffic patterns may be explored simultaneously, allowing for the quick assessment of various optimization and design options. Additionally, QC simulators are capable of simulating large-scale networks that are inaccessible to traditional computers \cite{daley2022practical}. The capacity to serve a large number of connected devices and a variety of applications with different performance needs is expected in next-generation networks, such as 6G, and this capability is critical for their design and optimization. In addition, QC simulations can offer important insights into how networks behave in harsh circumstances, like during cyberattacks or natural disasters, which can help with the creation of durable and adaptable 6G network architectures.

\textit{Intelligent Networks:} \textcolor{black}{The Q-RL models can  be integrated with federated learning architectures across distributed RAN and edge nodes, where local learning experiences are aggregated to improve global congestion control policies—without violating latency or privacy constraints. In operational terms, these Q-RL-based digital twins may reside in the non-real-time RAN Intelligent Controller layer of an O-RAN-compliant 6G architecture, where they continuously assess traffic trends and update control policies. When a predicted congestion state reaches a predefined threshold, the corresponding policy can be pushed to the near-real-time RIC or edge controller, enabling sub-second reconfiguration of routing tables, traffic shaping rules, or scheduling priorities—well before user experience degrades \cite{georgescu2014quantum}. } The incorporation of QC into network modeling and simulation could completely alter the way that communication networks are created, optimized, and run. By providing a more accurate and efficient way to explore complex network scenarios, QC can enable the development of more resilient, scalable and intelligent networks that meet the growing demands of 6G networks \cite{nawaz2019quantum}.

\subsection{The Path to the QC at the 6G Edge}

The convergence of quantum technologies and 6G networks has the potential to unlock a new realm of possibilities, revolutionizing the capabilities of specific applications and services  at the edge \cite{wang2022quantum, passian2022concept}. Quantum computing's  computational power and  security can provide solutions to complex problems, enhance user experiences, and drive innovation in 6G networks \cite{mastroianni2023quantum}. In the context of autonomous driving, vehicles generate massive amounts of data from sensors, cameras, and other onboard systems. Processing this data in real-time is crucial for making split-second decisions such as obstacle avoidance, path planning, and traffic management. \ac{QC} can enhance these processes by leveraging quantum algorithms that solve optimization problems such as transport route selection for many autonomous vehicles \cite{wang2021shaping}. For example,  \ac{QC} can optimize routes in real-time by rapidly calculating the most efficient path among countless possibilities, considering factors like traffic, road conditions, and potential hazards \cite{harwood2021formulating}. \ac{QC} algorithms can process sensor data, environmental information, and traffic patterns in real-time, enabling faster and more accurate decision-making for autonomous vehicles. This can significantly improve safety, efficiency, and traffic management \cite{mastroianni2023quantum}.

The high-fidelity rendering and complex interactions demanded by  \ac{XR} applications can benefit from \ac{QC}'s computational capabilities. \ac{QC} can accelerate complex simulations, enhance real-time rendering, and enable more realistic and immersive \ac{XR} experiences \cite{kamdjou2024resource}. The metaverse is a collective virtual shared space, integrating  \ac{VR}, \ac{AR}, and other digital environments—requires vast computational power to support real-time interactions, high-fidelity graphics, and complex simulations \cite{gencc2020design}. Integrating \ac{QC}  with mobile edge networks can address many of the challenges associated with delivering these rich, interactive experiences by providing enhanced computational capabilities right at the network’s edge. In the metaverse, \ac{MEC} reduces latency by processing data near the user, minimizing the delays that occur when data is sent back and forth to centralized data centers \cite{dong2022quantum}. \ac{QC} can significantly amplify this by enabling ultra-fast data processing and complex computations that classical computers struggle with. For instance, quantum algorithms can optimize the rendering of 3D environments, reducing the time required to generate and display complex scenes in \ac{VR} and \ac{AR}, leading to smoother, more realistic virtual experiences \cite{ai2024identifier}. \ac{QC}'s efficient compression and transmission algorithms can facilitate the streaming of high-quality volumetric video, enabling more immersive and interactive communication, entertainment experiences and healthcare applications \cite{ye2023smart}.

In the 6G network edge, public safety services such as predictive policing can be provided by the mobile network operators. \ac{QC} can be used to analyse historical crime data, social media, environmental conditions, and other inputs are examined to forecast possible criminal behaviors. \ac{QC} can easily handle the volume and complexity of this data, providing deeper insights and more precise forecasts than traditional computer approaches, which frequently struggle with it \cite{abuarqoub2021impact}. Real-time monitoring is another area where quantum computing can create a big difference when compared to the current monitoring systems. Video feeds from security cameras, drones, and other sensors can be processed by 6G edge with \ac{QC} to quickly identify known suspects or spot questionable activity \cite{amin2023detection}. This capacity improves situational awareness and makes it possible for law enforcement to react to new threats faster. Additionally, quantum algorithms can improve facial recognition and anomaly detection, making these systems more reliable and less prone to false positives, which is critical in high-stakes environments \cite{michail2023quantum}.

\ac{QC} at the edge has the potential to transform autonomous vehicle operations, logistics, and traffic management in the transportation sector. Urban traffic management is a challenging undertaking that requires in-the-moment traffic flow analysis, signal timing, and incident response. By swiftly evaluating large datasets, \ac{QC} can optimize these procedures and determine the most effective routing routes, dynamically modify traffic lights, and anticipate traffic jams before they occur \cite{cooper2021exploring}. As a result, there is less congestion, shorter travel times, and less pollutants. Additionally, supply chain management and logistics depend heavily on \ac{QC}. By evaluating real-time data on vehicle locations, weather, and road conditions, Edge \ac{QC} can help optimize delivery routes, minimize fuel usage, and shorten delivery times \cite{pfister2024transfer}. Quantum algorithms excel at solving these complex optimization problems, which are often beyond the capabilities of classical computers, particularly when multiple variables must be considered simultaneously.

\subsection{6G Quantum Transport Networks}

As mobile networks evolve, the demand for higher efficiency, lower latency, and enhanced security becomes increasingly critical where the quantum technologies  can impact 6G transport networks. For example, long-distance signal amplification is possible with amplifiers in classical communication. Nevertheless, direct amplification of quantum signals—which entail quantum states like superposition and entanglement—is prohibited by the \textit{no-cloning theorem}, which states that quantum information cannot be properly replicated without disturbing the state. This poses a serious obstacle to the long-distance quantum information transmission needed for 6G network connection. Quantum repeaters, which offer a way to maintain the integrity of quantum signals over great distances, present an answer to this issue. Quantum repeaters work by dividing a communication path into smaller segments and placing repeaters in between \cite{azuma2023quantum}. They make use of a technique known as entanglement swapping, in which different points along the communication line generate pairs of entangled particles.  Without directly sending the quantum states over the entire distance, the entanglement can be gradually extended throughout the entire path by carrying out quantum measurements at each repeater node \cite{pu2021experimental}. This maintains the security and characteristics of the quantum information by enabling the reconstruction of the original quantum state at the endpoint. To further improve communication dependability, quantum repeaters also use error correction techniques to adjust for any noise or flaws produced during transmission. Large-scale quantum networks, where data security is critical, including the quantum internet and secure 6G transport systems, depend on quantum repeaters \cite{van2013designing}.   The quantum teleportation technique uses the laws of quantum physics to transfer quantum information—like the state of a qubit—between two points without requiring the actual movement of the particles involved \cite{bebrov2020teleportation}. This technique is particularly groundbreaking for 6G transport networks since it permits instantaneous and secure information transfer independent of the physical distance between the communication parties. Quantum teleportation transmits a qubit's state via entanglement as opposed to conventional data transfer, which uses optical fibers or other media \cite{hu2023progress}. This eliminates the need for direct physical transmission and significantly improves security.

Quantum-enhanced signal processing is poised to revolutionize 6G networks by leveraging the concepts of quantum mechanics to increase data processing and transmission performance and efficiency. With 6G networks expected to offer unprecedented data throughput, ultra-low latency, and very reliable communication, quantum-enhanced methods can be very beneficial.  Quantum-enhanced signal processing in 6G networks can use quantum technology and algorithms to carry out operations like interference control, noise reduction, and signal recognition more accurately than with conventional techniques \cite{blais2020quantum}. Quantum algorithms, for instance, may examine intricate signal patterns and identify irregularities or interference with greater sensitivity, improving the performance and dependability of 6G transport networks \cite{awschalom2021development}. This capability is crucial for managing the high data rates in 6G networks, where traditional signal processing methods may struggle to keep up. One key aspect of quantum-enhanced signal processing is its ability to handle quantum states of light or other particles used in 6G communication infrastructure. With the use of quantum qualities like superposition and entanglement, quantum sensors are able to measure signal parameters like phase and amplitude with greater accuracy \cite{guo2020experimental}.  The overall characteristics of quantum technologies for the 6G core, edge and transport networks with the related challenges is given in Table \ref{table3}.

\begin{table*} [htp!] 
\centering
\captionsetup{font=sc}
\scriptsize
\caption{Quantum Technology based Applications for 6G Core, Edge and Transport Networks}
\begin{tabular}{|p{1cm}|p{3.5cm}|p{4cm}|p{4cm}|p{2.3cm}|}
\hline
\rowcolor{gray!30}
\hspace{.2cm}\textbf{6G} & 
\hspace{.9cm}\textbf{6G Application} & 
\hspace{.9cm}\textbf{Potential Benefits} & \hspace{.8cm}\textbf{Potential Challenges}  & \hspace{.5cm}\textbf{References}\\ \hline
& Traffic Management             & Real-time optimization, reduced latency, improved user experience & Scalability, computational power, integration with existing systems  &  \cite{li2020quantum,bhatia2020quantum} \\ \cline{2-5}
& Congestion Control              & Proactive congestion prevention, efficient resource allocation & Algorithm development, real-time adaptation &  \cite{qu2022temporal} \\ \cline{2-5}
& Spectrum Management            & Dynamic spectrum allocation, enhanced spectral efficiency & Interference management, regulatory frameworks & \cite{alsaedi2023spectrum,liu2016optimization,zhang2022quantum}  \\ \cline{2-5}
Core & Fault Tolerance                & Improved network reliability and resilience & Hardware scalability, error correction  & \cite{liu2023enabling,egan2021fault} \\ \cline{2-5}
& Simulation and Modeling         & Accurate network modeling, efficient design and optimization & Computational complexity, model validation  & \cite{zhou2020limits,daley2022practical}  \\ \cline{2-5}
& Intelligent Networks         &  Development of intelligent and self-learning networks & Implementation aspects, model generation  & \cite{georgescu2014quantum,nawaz2019quantum}  \\ \hline
& Autonomous Driving             & Real-time data processing, improved decision-making for safety & Sensor integration, ethical considerations & \cite{passian2022concept,mastroianni2023quantum,wang2021shaping} \\ \cline{2-5}
& Extended Reality  & Enhanced rendering, realistic and immersive experiences & Hardware limitations, power consumption & \cite{kamdjou2024resource,gencc2020design} \\ \cline{2-5}
& Volumetric Video Streaming     & Efficient compression and transmission, high-quality streaming & Bandwidth requirements, storage limitations & \cite{dong2022quantum,ai2024identifier,ye2023smart} \\ \cline{2-5}
Edge  & Metaverse   & Secure and privacy-preserving interactions & Quantum-resistant cryptography, user authentication &  \cite{khalid2023quantum,chehimi2023quantum} \\ \cline{2-5}
& Generative AI Assistants       & Contextually aware, nuanced interactions & Training data bias, ethical concerns &  \cite{li2020quantum,peral2024comparing,romero2021variational}  \\ \cline{2-5}
& Digital Twin                   & Real-time synchronization, accurate simulations & Data security, computational complexity &  \cite{lv2022digital,paul2024digital,bariah2023digital} \\ \cline{2-5}
& Law Enforcement and Transportation & Enhanced surveillance, real-time data analysis & Privacy concerns, ethical considerations &  \cite{abuarqoub2021impact,amin2023detection,michail2023quantum} \\ \cline{2-5}
 & Edge Intelligence Services     & Optimized pricing strategies, efficient resource allocation & Cost-effectiveness, demand prediction & \cite{wang2022quantum,harwood2021formulating,yarkoni2020quantum,cooper2021exploring,pfister2024transfer} \\ \hline
 & Quantum repeaters & Realization of a global quantum communication networks & Quantum state preservation, synchronization and timing issues &  \cite{azuma2023quantum,pu2021experimental,van2013designing} \\ \cline{2-5}
 Transport &  Quantum teleportation based communication  & Instantaneous communication of quantum states over large distances &  Entanglement distribution, quantum measurement  &   \cite{bebrov2020teleportation,hu2023progress} \\ \cline{2-5}
 & Quantum-enhanced signal processing &  Analyze, and interpret signals, with far-reaching implications & Noise sensitivity, error Rates and fault tolerance & \cite{blais2020quantum,awschalom2021development,guo2020experimental} \\ \hline
 \end{tabular} 
\label{table3}
\vspace{-0.5cm}
\end{table*}

\section{Quantum Technologies  for  6G Security and Privacy}
\label{sec:security}



\textcolor{black}{Security is one of the most critical aspects of 6G, and quantum technologies offer promising solutions. 
Quantum technologies can play a vital role in enhancing the security of 6G networks and applications. As 6G networks become more complex and interconnected, ensuring robust security measures will be crucial to protect sensitive data and communication. In this section, some ways that quantum technologies can contribute to 6G security are described and discussed next.
}

\subsection{Quantum Key Distribution}

\ac{QKD} is a secure communication method that uses a cryptographic protocol that incorporates components of quantum mechanics. Quantum key distribution can provide secure and unbreakable encryption for communication channels in 6G networks. \ac{QKD} is an advanced cryptographic method that leverages quantum mechanics to guarantee secure communication \cite{sharma2021quantum}. In contrast to classical encryption methods, which rely on the computational difficulty of certain mathematical problems, \ac{QKD} offers a theoretically unbreakable security model based on the fundamental laws of quantum physics. The essence of \ac{QKD} lies in its capability to securely transfer encryption keys between two parties, with any eavesdropping attempts being detectable. This is accomplished using quantum bits (qubits), which can exist in multiple states at once due to the principle of superposition. When a qubit is observed or measured, it collapses into a specific state. This process is inherently detectable, so the communicating parties know if their key exchange has been compromised. The architecture of a \ac{QKD} network consists of several key components that work together to enable the secure generation, distribution, and management of quantum keys.  The steps involved for \ac{QKD} are as follows \cite{cao2022evolution}.

\begin{itemize}[leftmargin=*]
    \item \textit{Quantum Transmitters and Receivers:} The quantum transmitter, referred to as Alice, sends quantum bits (qubits) to the receiver. These qubits are typically encoded in the polarization or phase of photons, which are transmitted over a quantum channel, such as optical fiber or free space. The quantum receiver (i.e., Bob) detects the qubits sent by Alice. Bob measures the incoming qubits to extract the key bits. Due to the principles of quantum mechanics, any eavesdropping attempt will interfere with the qubits and can be detected by Bob.
    \item \textit{Quantum Channel:} The quantum channel is the medium through which the qubits are transmitted. This can be an optical fiber connection or a free-space optical link. The quantum channel is very sensitive to noise and losses, which are critical factors for the performance and range of the QKD system.
    \item \textit{Classical Channel:} Alongside the quantum channel, a classical communication channel is used for exchanging information between Alice and Bob. This channel is essential for performing key sifting, error correction, and privacy amplification processes. Unlike the quantum channel, the classical channel does not require quantum security but should still be protected by conventional encryption methods.
    \item \textit{Key Sifting:} After Bob measures the qubits, Alice and Bob perform a key sifting process over the classical channel. This step involves comparing a subset of the transmitted and received qubits to identify and discard any bits that may have been altered or lost during transmission due to either noise or potential eavesdropping.
    \item \textit{Error Correction:} The key obtained after sifting may still contain errors due to imperfections in the quantum channel. Error correction algorithms are applied to reconcile the differences between Alice's and Bob's keys, ensuring that both parties have identical keys.
    \item \textit{Privacy Amplification:} Privacy amplification is a post-processing step where Alice and Bob shorten their keys to eliminate any partial information that an eavesdropper might have gained. This process results in a final secret key that is highly secure and ready for cryptographic use.
    \item \textit{Key Management System (KMS):} The Key Management System is responsible for managing the generated quantum keys. It handles key storage, distribution, and integration with cryptographic applications. The KMS ensures that the keys are securely delivered to the intended devices or applications.
    \item \textit{QKD Network Control and Monitoring:} The QKD network requires a control and monitoring system to manage the quantum and classical channels, monitor the performance of the QKD devices, and detect any anomalies or security breaches in real-time.
\end{itemize}

\subsubsection{Integration of QKD in B5G/6G Networks}
\begin{figure}[htp!]
    \centering
    \includegraphics[width=\linewidth]{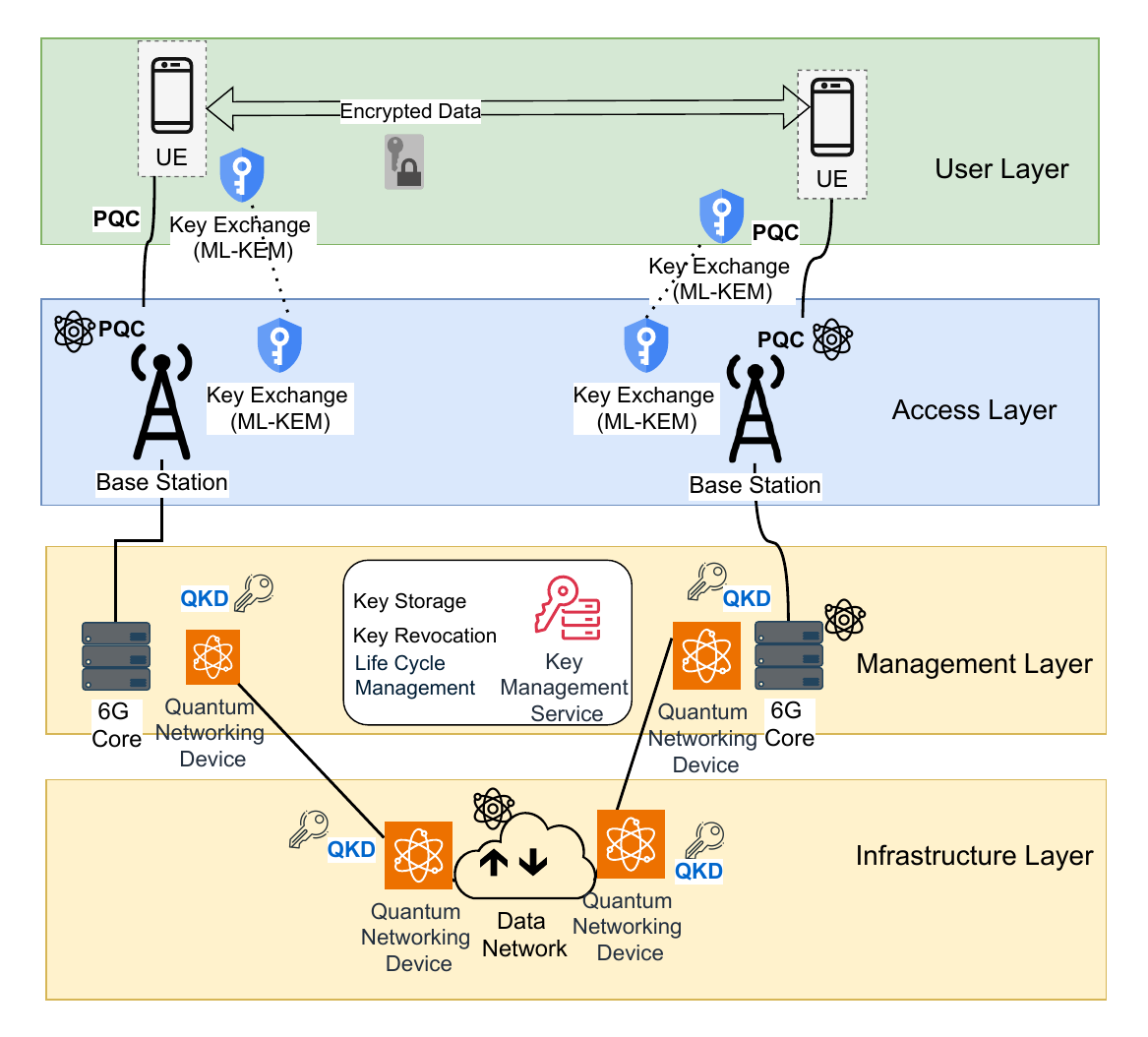}
    \caption{Potential QKD and PQC Placements within 6G.}
    \label{fig:qkdin6g}
    \vspace{-0.4cm}
\end{figure}

The potential placements of QKD have been displayed in Fig. \ref{fig:qkdin6g}. In this subsection, we describe the importance of such a quantum technology addition to the 6G networks.

\textit{Role of QKD in B5G/6G Networks:} On the way to B5G and 6G networks, securing communication channels is becoming increasingly important. Traditional encryption methods that are still effective today may become vulnerable as computing power increases, especially with the advent of quantum computers capable of breaking current cryptographic algorithms. At this point, QKD becomes indispensable as it offers unattainable security with traditional encryption methods \cite{fujiwara2016unbreakable}. QKD makes it virtually impossible for attackers to decrypt the communication, even with the most powerful classical or quantum computers \cite{grover2024security}. Thus, QKD integration becomes a critical component in the security infrastructure of B5G/6G networks, ensuring that they remain secure even in the face of future technological advancements with its secure key exchange approach \cite{james2023key}. 

\textit{Importance of QKD in B5G/6G Security:} The integration of QKD into B5G/6G networks is not just a theoretical exercise but a practical necessity for several reasons. In future networks, the volume and sensitivity of transmitted data will increase significantly. QKD's ability to detect any form of eavesdropping on key exchanges makes it an indispensable tool for protecting this data \cite{kamran2021evaluation}. B5G/6G networks will play a crucial role in supporting critical infrastructure, such as smart grids, autonomous vehicles, and healthcare systems. The security of these systems is of paramount importance, and QKD provides a level of security that classical cryptographic methods cannot match \cite{lopez2019applying}. By securely distributing keys, QKD helps to protect these important systems from cyber-attacks and espionage. As data protection regulations around the world become increasingly stringent, the need for secure communication methods that can ensure the protection of sensitive information is growing  \cite{van2022towards}. QKD, with its unparalleled security guarantees, can position B5G/6G networks to effectively meet these regulatory requirements and help avoid potential legal and financial repercussions of data breaches. The success of B5G/6G networks will depend to a large extent on the trust that users and industry place in them \cite{bajric2023enabling}. This trust is an essential prerequisite for the broad acceptance and success of future networks.

\subsubsection{Potential Issues}

QKD plays a critical role in improving the security of B5G/6G networks by providing a robust, future-proof solution for secure cryptographic key exchange. Integrating QKD into existing security frameworks such as Virtual Private Networks (VPNs) based on IPsec and MACsec enhances the security of these protocols by using quantum-generated keys for encryption \cite{niemiec2016authentication}. QKD can be integrated into these frameworks as follows:

\begin{itemize}[leftmargin=*]
    \item \textit{QKD-Enhanced IPsec (Quantum IPsec):} IPsec (Internet Protocol Security) is a set of protocols designed to ensure secure communication across IP networks \cite{oppliger1998security}. It uses cryptographic algorithms to ensure confidentiality, integrity, and authentication of data. QKD can be integrated with IPsec by replacing the traditional key exchange mechanisms (such as IKE - Internet Key Exchange) with quantum-generated keys \cite{marksteiner2014approach}. The QKD system continuously generates and supplies fresh keys to the IPsec system, ensuring that each session key used in the encryption process is secure and resistant to attacks, including those from quantum computers. The QKD system can be connected to the IPsec KMS to provide quantum keys directly for IPsec sessions. The IPsec protocol would be modified to allow for the seamless insertion of quantum-generated keys into the IPsec Security Associations (SAs), where the keys are used for encrypting and authenticating IP packets.
    
    \item  \textit{QKD-Enhanced MACsec (Quantum MACsec):} MACsec (Media Access Control Security) is a layer two security protocol that provides encryption, integrity, and data origin authenticity for Ethernet frames \cite{dik2023open}. Similar to IPsec, MACsec can benefit from QKD by using quantum-generated keys for the encryption and integrity protection of Ethernet frames at the B5G/6G Core. The keys generated by the QKD system would replace those traditionally generated by MACsec's Key Agreement Protocol (MKA) \cite{lyssenko2023leveraging}. The QKD system would distribute quantum keys to MACsec-enabled devices through a secure management interface. The QKD-generated keys would then be used to establish secure MACsec sessions between network devices, ensuring that each frame transmitted over the network is protected with quantum-resistant encryption \cite{cho2021using}.
\end{itemize}

By integrating QKD, both IPsec and MACsec can benefit from the unbreakable security of quantum keys, which secure the VPNs against current and future cryptographic attacks. QKD enables more frequent key refresh rates, improving security by limiting the exposure of a particular key. This is especially important in environments that require a high level of security. The integration of QKD prepares IPsec and MACsec VPNs for the post-quantum era and ensures that they remain secure against attacks from quantum computers \cite{gazdag2023quantum}. 

\subsection{Quantum Secure Direct Communication}

\ac{QSDC} is a secure communication technique that utilizes quantum mechanics to transmit information directly between parties, eliminating the need for encryption or traditional key exchange \cite{pan2024evolution}. Unlike \ac{QKD}, where a key is shared and then used to encrypt the message, \ac{QSDC} sends the message itself in a secure quantum channel, ensuring that any attempt to eavesdrop can be detected. The security of \ac{QSDC} is based on the no-cloning theorem and quantum superposition, meaning that any interception of the quantum information alters the state of the transmitted qubits, alerting the communicating parties to the presence of an eavesdropper. This makes \ac{QSDC} ideal for transmitting highly sensitive information with the assurance that it cannot be intercepted without detection \cite{long2007quantum}. \ac{QSDC} could play a pivotal role in enhancing the security of 6G networks, particularly for applications where ultra-secure, low-latency communication is critical \cite{dabas202426g}. Since 6G networks are anticipated to support vital infrastructure like smart grids, autonomous transportation, and healthcare systems, \ac{QSDC} could offer quantum-level security for these communication channels, ensuring that sensitive data remains protected from interception or tampering without detection. Furthermore, 6G networks are expected to be a stepping stone towards the quantum internet. \ac{QSDC} could be a fundamental component in securely transmitting quantum information, creating a secure foundation for future quantum communication systems integrated with 6G. For government and military applications, \ac{QSDC} within 6G networks can provide a secure communication layer for command and control operations, intelligence sharing, and diplomatic communications, ensuring that these critical communications are impervious to interception.


\subsection{Quantum Error Correction for 6G Reliability}

One of the key challenges for incorporating quantum techniques in 6G networks is ensuring reliability, especially given the fragility of quantum states. \ac{QEC} techniques are essential to overcome the inherent noise and errors in quantum systems, providing a robust foundation for quantum-enabled 6G infrastructure \cite{lidar2013quantum}. Quantum systems are highly susceptible to various types of noise, including decoherence, photon loss, and operational errors, which can lead to information loss and transmission errors \cite{paladino20141}. Unlike classical errors, which involve flipping or altering bits, quantum errors can involve both bit-flip and phase-flip errors due to the superposition and entanglement properties of quantum states. In the context of 6G, which aims to achieve \ac{uRLLC}, managing quantum noise is critical for maintaining the integrity and reliability of data transmission in quantum-enhanced communication systems \cite{pradhan2024survey}. \glspl{QEC} address this problem by encoding quantum information into larger quantum systems in such a way that errors can be detected and corrected without directly measuring the quantum state \cite{knill1997theory}. This prevents the destruction of quantum information, allowing for reliable communication even in the presence of quantum noise. Techniques such as the Shor code, Steane code, and surface code are examples of quantum error correction mechanisms that can be adapted for 6G quantum communications \cite{brun2019quantum}.

Incorporating quantum error correction into 6G networks involves several challenges and opportunities: \textit{(1) Hybrid Quantum-Classical Error Correction:} Given the early stages of quantum technology, it is expected that 6G will initially rely on hybrid quantum-classical systems. Quantum data will be transmitted alongside classical data, and \ac{QEC} will work in conjunction with classical error correction methods (like low-density parity-check codes) to ensure overall network reliability. This hybrid approach will help in managing errors across both classical and quantum communication layers \cite{endo2021hybrid}. \textit{(2) Entanglement Distribution and Error Correction:} A key feature of future 6G quantum networks is the distribution of entangled states across long distances. For long-distance quantum communication, quantum repeaters will be essential to extend the range by applying error correction to entangled states, thereby preserving entanglement across noisy channels. The application of quantum error correction codes within repeaters can maintain entanglement fidelity and improve the overall reliability of quantum links \cite{dur2007entanglement}.
\textit{(3) Fault Tolerant Quantum Communication:} As 6G evolves to accommodate quantum computing and communication, fault-tolerant quantum error correction will be crucial. This requires encoding quantum information in a way that allows for the correction of multiple errors without decoding the quantum state, providing reliable communication even in noisy environments. Fault-tolerant schemes like the surface code can correct errors caused by noise while allowing scalable quantum operations, making them ideal for 6G quantum networks \cite{childress2006fault}.

However, in order to achieve robust and scalable quantum error correction for 6G, several future research directions must be explored \cite{nawaz2019quantum}. These include optimizing \glspl{QEC} algorithms for real-time applications, integrating quantum repeaters with \ac{QEC} to extend communication range, and developing hardware-efficient \ac{QEC} methods compatible with 6G’s stringent latency and reliability requirements. Additionally, implementing machine learning algorithms to predict and correct errors in quantum systems presents a promising avenue for enhancing the performance of \ac{QEC} in practical quantum communications \cite{zoratti2023improving}.

\subsection{Quantum-resistant Cryptography}

As 6G networks emerge, providing faster data rates, lower latency, and extensive device connectivity, security remains a paramount concern. The advancement of quantum computing poses a direct threat to traditional cryptographic protocols that secure modern communication networks. This risk necessitates the integration of \ac{PQC} into the security architecture of 6G to ensure resistance against quantum-enabled attacks. The cryptographic technique most commonly used in today's telecommunication networks is asymmetric cryptography, which is based on the computational complexity of mathematical problems \cite{mehic2023quantum}. However, existing asymmetric cryptographic standards like RSA and ECC, which depend on the computational difficulty of factoring large numbers and solving discrete logarithms, are susceptible to quantum attacks. Quantum algorithms like Shor’s algorithm can exponentially reduce the time required to break these cryptographic systems, making classical encryption and key exchange schemes obsolete in the post-quantum era.   The integration of \ac{PQC} into telecommunication frameworks is essential for strengthening security against emerging quantum threats. 6G security protocols could integrate \ac{PQC}, which utilizes mathematical problems that are difficult for both classical and quantum computers to solve, providing long-term security against quantum attacks \cite{9727214}.

Table \ref{tab:comprehensive_comp} provides a comprehensive comparison of quantum security technologies for 6G.

\begin{table*}[htp!]
\captionsetup{font=sc}
\centering
\scriptsize
\caption{Comprehensive Comparison of Quantum Security Technologies for 6G 
}
\label{tab:comprehensive_comp}
\begin{tabular}{|p{2.7cm}|p{3cm}|p{3cm}|p{3cm}|p{4cm}|}
\hline
\rowcolor{gray!30}
\textbf{Feature}                     & \textbf{Quantum Key Distribution (QKD)}   & \textbf{Quantum Secure Direct Communication (QSDC)} & \textbf{Post-Quantum Cryptography (PQC)}     & \textbf{Quantum Error Correction for 6G Reliability}        \\ \hline
\textbf{Description}                 & Secure key exchange via quantum states  \cite{sharma2021quantum}  & Direct communication without encryption   \cite{pan2024evolution}          & Classical algorithms resistant to quantum attacks \cite{9727214} & Detection of communication data errors in quantum channels \cite{pradhan2024survey}  \\ \hline
\textbf{Advantages}                  & Theoretically unbreakable encryption      & No need for key distribution                        & Works with existing infrastructure              & Highly reliable communications     \\ \hline
\textbf{Current Limitations}         & Limited range and high cost               & Complex hardware requirements                       & High computational load                         & Scalability issues, high processing power     \\ \hline
\textbf{Key 6G Use Cases}            & Financial sector, government communication & Military, critical infrastructure                   & General-purpose encryption for 6G               & General-purpose error detection in 6G Core        \\ \hline
\textbf{Technology Readiness Level (TRL)} & TRL 6 – Demonstrated in relevant environments & TRL 5 – Early testing phase                        & TRL 7 – Mostly ready for deployment             & TRL 4 – Research in progress                 \\ \hline
\end{tabular}
\vspace{-0.5cm}
\end{table*}

\color{black}

\section{Deployment Challenges}
\label{sec:deployment}



\begin{figure}[htp!]
    \centering
    \includegraphics[width=0.85\linewidth]{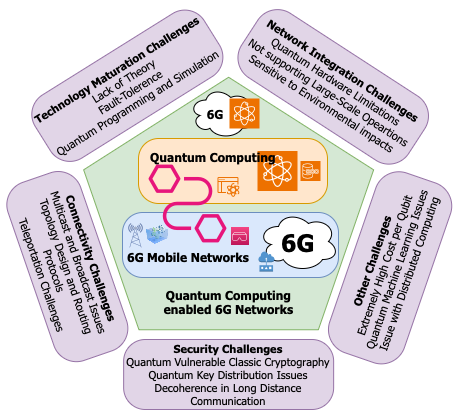}
    \caption{Quantum computing deployment challenges in 6G.}
    \label{fig:Challenges}
    \vspace{-0.5cm}
\end{figure}

\textcolor{black}{The revolutionary convergence of emerging 6G networks with quantum computing is envisaged to underpin next generation network operation, optimization and applications. However, being a maturing technology, a number of challenges need to be addressed towards realizing quantum computing enabled 6G networks~\cite{yang2023survey}. This section will explore the challenges under technology maturation challenges, integration challenges, connectivity challenges, security challenges, and other challenges, as illustrated in Fig.~\ref{fig:Challenges}, and explore possible solutions.}

\subsection{Quantum Computing Technology Maturation Challenges}
\subsubsection{Challenges}

Science has been advancing rapidly towards realizing quantum computers. However, not every concept in quantum computing is understood fully. For instance,  quantum entanglement is a concepts that is yet to be fully explored as the complete understanding of the properties relating to the correlation of entangled qbits is yet to be unraveled~\cite{yang2023survey}. Hence, entanglement is understood through mathematical formulations of quantum mechanics. This demands quantum computing related knowledge to mature before using quantum computing as a fully fledged technology for 6G communication networks and applications. Furthermore, quantum computing hardware is identified to be in the \ac{NISQ} state. This signifies that quantum computers are very well capable of performing non-trivial quantum computations. However, \ac{NISQ} hardware still needs developments to be fault-tolerant and achieve the full advantages of quantum computing. Hence, environmental conditions can act as noise and cause the loss of coherence within quantum states. It should be noted that 6G communication network and infrastructure should be fault-tolerant and operate in noisy environments to delay-sensitive time-critical real-time applications~\cite{de2021survey}. This demands the maturation of quantum computing fault-tolerance techniques and noise resistant quantum computing hardware towards revolutionising 6G and beyond communication networks. In addition, being a revolutionary computing paradigm, programming quantum computers is unique compared to classical computing models. This demands novel approached to program quantum computers. For instance, Qiskit is a software stack developed for quantum computing by IBM~\cite{qiskit}. Qiskit can be used to build programs to be executed in IBM and third-party quantum computing hardware. However, considering the way quantum computers work, where measuring and restoring a state in a quantum computer is not possible, debuggers running in actual quantum computing hardware are yet to be developed~\cite{yang2023survey}. Thus, testing a quantum program requires it to be run in real quantum computer. Hence developing sophisticated programs for quantum computing demands running test codes in actual hardware a large number of times, which is a challenge. In addition, there is no solid approach of integrating noise in existing quantum simulators. Therefore, quantum computing technology needs maturation to have better testing and development tools to become an integral component of emerging 6G networks, especially considering the sophisticated nature of 6G networks and applications.

\subsubsection{Possible Solutions}

Even though the concept of quantum entanglement has not yet been researched, it has been proven by experiments. Moreover, quantum mechanics is able to predict quantum computer systems~\cite{yang2023survey}. In addition, quantum computing research is exploring the evolution of \ac{NISQ} towards \ac{PISQ}~\cite{bertels2021quantum}. \ac{PISQ} proposes the use of what is already known about qubits to define algorithms and develop applications. Furthermore, the development of solid quantum programming languages and debuggers is an active area of research towards enabling quantum computing as a promising solution for the 6G era. For example, Cirquo, a unit testing approach and debugging method for quantum computing, is proposed in ~\cite{metwalli2024testing}. These developments strengthen quantum computing as a mature technology for the beyond 5G and 6G wireless communication networks.

\subsection{Integration Challenges of Quantum Computers with 6G}
\subsubsection{Challenges}

The advancements communication networks towards 6G envisages to integrate heterogeneous hardware and networks towards creating a seamless weave of connected devices. This demands computing resources to extend from large server farms to network edge devices~\cite{de2021survey}. Conversely, quantum computing hardware comes with their own unique requirements. For instance, the preservation of quantum information demands isolating quantum states from environmental impacts~\cite{yang2023survey}. Hence, quantum computing operations need to be performed ensuring that quantum states are not drifted to preserve the magnitude of qubits to prevent accumulating errors. Moreover, existing quantum hardware are not capable of performing large-scale operations.These issue should be addressed towards integrating quantum computing hardware with 6G network infrastructure.  Furthermore, quantum processor hardware requires their edge to be routed with complex wiring, making the packaging of qubits, which is the process to control and house quantum hardware containing qubits, a significant challenge towards realising quantum computers. This challenges the integration of quantum hardware with heterogeneous 6G nodes and devices, demanding streamlining qubit packaging process and technologies~\cite{100quibit}. In addition, the fully functional quantum computers are very large and are reminiscent of early computers~\cite{yang2023survey}. This is because the operation of quantum computers requires special environmental conditions, which means that the processing environment is large despite the compact size of the quantum processors. For example, a vacuum chamber is needed to maintain quantum states, portals to the chamber are needed for optical inputs, optical sources are needed for optical operations, and cooling materials are needed to control the temperature of the quantum computer, which takes up a large space.

\subsubsection{Possible Solutions}

Despite the existing challenges, research work advances towards addressing challenges of integrating quantum computing with 6G network infrastructure. For instance, a quantum error correction utilising the surface code is presented in~\cite{krinner2022realizing}. In addition, a stable error-corrected logical qubit is demonstrated in~\cite{sivak2023real}. These work demonstrate efficient and high performance quantum error correction cycles towards realising error free computation in 6G network infrastructure. Moreover, technologies such as multi-level wiring indicate promising prospects of efficient packaging of qubits towards a future with quantum computing integrated 6G networks~\cite{100quibit}. This would also facilitate quantum computers be more space efficient, high quality and high performance (better Circuit Layer Operations Per Second - CLOPS), enabling them to power 6G infrastructure. In addition, technologies, devices and applications of quantum computing demands novel standardisation that would facilitate the interoperability between platforms, vendors, and applications~\cite{mazurquantum}.

\subsection{Connectivity Challenges of QC with 6G}
\subsubsection{Challenges}

Despite being identified as a powerful enabler of 6G an beyond networks, quantum computing also comes with a number of connectivity challenges. For instance, establishing end-to-end communication in a quantum network has several challenges~\cite{yang2023survey}, including transmitting and receiving data with multiple nodes simultaneously, and broadcasting data in a quantum network. Moreover, quantum network topology design and quantum routing protocols are still in a premature state, making it another connectivity challenge related to deploying quantum computing in 6G networks.

Furthermore, as quantum states cannot be cloned, it is a challenge to store and transmit quantum data. This also leads to not being able to generate redundancies or re-transmit data. This means any data that is lost can be lost forever. As quantum networks operate on teleportation, which is based on entanglement, teleportation related challenges should also be resolved towards realising quantum based 6G networks. For instance, any data that is lost during teleportation cannot be recovered. This makes it challenging for quantum computing enabled networks to reach the strict quality of standards in emerging networks. Furthermore, it is a challenge to communicate across long distances due to decoherence. This is due to quantum states losing information during transmission. The utilisation of quantum repeaters are explored to facilitate long distance transmission of quantum states. Yet, it is a challenge due to the difficulty of cloning quantum states~\cite{azuma2023quantum}.

\subsubsection{Possible Solutions}

Quantum computing research aims to address connectivity challenges to integrate quantum computers with future 6G communication networks. For instance, an entanglement routing model and an entanglement routing algorithm is presented in~\cite{shi2020concurrent} to solve the entanglement routing problem. Furthermore, the utilisation of quantum multiplexers are explored to facilitate routing in quantum networks~\cite{yang2023survey}. Moreover, the utilisation of quantum switches has been thoroughly explored in network typologies such as the star network topology. In addition, research is exploring novel methods to enhance quantum teleportation fidelity~\cite{guo2020fidelity}. Furthermore, researchers are exploring reliable communication of quantum states across optical, wireless channels, satellite and underwater channels~\cite{hasan2023quantum}. These technological developments also demand solid standardisation approaches to allocate spectrum and streamline the advancements of quantum computing.

\subsection{Security Challenges of Deploying QC with 6G}
\subsubsection{Challenges}

Advancements of communication networks towards 6G expand attack surfaces and demands more attention to strengthen security~\cite{de2023survey}. Securing a network requires ensuring confidentiality, integrity, and availability. This has become more challenging with the developments of quantum computing as an enabler of 6G and beyond networks. For instance, the security of communication networks depend on cryptography and the advents of quantum computing has endangered many classical cryptography approaches. This demands exploring quantum resistant post-quantum cryptography. This challenge is further intensified due to the increasing utilisation of cryptography applications such as blockchain in 6G and beyond networks~\cite{de2021survey}.  \ac{QKD} is being developed as a solution to provide symmetric key encryption for post-quantum cryptography~\cite{cao2022evolution}. However, \ac{QKD} has yet to mature into a full-fledged key distribution technology and requires standardisation so that it can be used by multiple parties to ensure security in post-quantum computing networks. \ac{QKD} also faces the challenge of noise interfering transmission. The interference from noise can be identified as an intrusion, resulting in a false positive. This would lead to \ac{DoS}, which would affect the availability of networks and services~\cite{yang2023survey}. In addition, the key efficiency of \ac{QKD} is low, demanding improved approaches to secure quantum enabled 6G and beyond networks. In similar vein, it is challenging to use the zero-trust network architecture in quantum networks as network nodes in quantum enabled networks are designed to trust each other. This leaves another challenge to secure quantum enabled networks in 6G and beyond era.

\subsubsection{Possible Solutions}

\ac{QKD} is being developed as a solid technology to secure networks in the post quantum era. For instance, \ac{QKD} is explored in~\cite{cao2022evolution} elaborating its potential, elements, interfaces, protocols, physical/network layers, standardisation requirements, and applications. Furthermore, \ac{QKD} can be integrated with other technologies and concepts such as network coding~\cite{de2012robust} to provide security and robustness in 6G and beyond networks. Mitigating noise in communication channels by utilising optical fibres with more photon quality can reduce the possibility of false positives in \ac{QKD}.  In addition to \ac{QKD}, classical cryptography techniques are also explored to be used in the post-quantum era. Unlike factorization-based cryptography techniques that are susceptible to quantum attacks, cryptography techniques such as lattice-based cryptography is identified to be quantum proof and can be used in the post quantum era~\cite{yang2023survey}. Not only \ac{QKD} and classical cryptography techniques, but also hybrid approaches between them are also explored to secure post-quantum 6G and beyond networks.

\subsection{Other Challenges of Deploying QC with 6G}
\subsubsection{Challenges}

In addition to technology maturation, integration, connectivity, and security challenges of deploying quantum computing in 6G and beyond networks, there are a number of other challenges that should be addressed. For instance, the cost of quantum computers are extremely high, making it a technology that cannot be used to operations in mobile networks at the moment. The cost per qubit presently ranges around \$10,000, whereas network operations require millions of qubits being processed~\cite{yang2023survey}. This is a significant challenge as network operators, service providers and application developers run as businesses that thrives to minimise costs. This demands, smaller, cost-efficient, noise-immune quantum computers to power 6G and beyond communication networks. \textcolor{black}{Another critical bottleneck for the practical deployment of quantum technologies for 6G networks lies in the current limitations of quantum hardware. The coherence time, i.e. the amount of time a quantum system can maintain its quantum state without decoherence, is still extremely short for many quantum devices, especially \ac{NISQ} systems. This limitation significantly affects the reliability of quantum operations in time-sensitive applications such as real-time RAN processing or low-latency core decisions. Moreover, scaling the number of qubits while maintaining fidelity is a formidable engineering challenge. Existing systems support tens to a few hundred qubits, but robust network operations and quantum-assisted AI tasks for 6G will likely require thousands of fault-tolerant qubits, requiring breakthroughs in quantum error correction and miniaturization of physical qubits. Furthermore, cryogenic integration remains a major challenge, as superconducting qubits and other leading-edge platforms require ultra- low temperatures (typically in the millikelvin range) to function. The integration of such systems into mobile or edge environments (e.g. base stations, smart city infrastructure) requires the development of compact, energy-efficient cryogenic cooling systems and chip-scale packaging techniques. Without these advances, the possibility of embedding quantum processors in distributed 6G infrastructures remains limited to highly specialized or centralized deployments (e.g. quantum-enabled data centers or satellites).}

\subsubsection{Possible Solutions}

The power of quantum computers need to be explored to handle existing machine learning techniques. Conversely, quantum machine learning has become an active area of research, where quantum annealing is being developed as a technique to find global minima of optimisation problems~\cite{rajak2023quantum}. In addition, game theory based approaches are explored to realise distributed quantum machine learning in 6G and beyond networks.

Finally, Table \ref{tab:comprehensive_comp} summarizes the deployment challenges and solutions for the use of quantum technologies in 6G.

\begin{table*}[htp!]
\captionsetup{font=sc}
\centering
\scriptsize
\color{black}
\caption{Deployment Challenges and Solutions for Quantum Technologies in 6G}
\label{tab:deployment_challenges}
\begin{tabular}{|p{1.8cm}|p{3.7cm}|p{3.5cm}|p{3.5cm}|p{3.5cm}|}
\hline
\rowcolor{gray!30}
\textbf{Feature}                     & \textbf{Technological Maturity}         & \textbf{6G Infrastructure Integration}                    & \textbf{Connectivity Challenges} & \textbf{Security Challenges}            \\ \hline
\textbf{Description}                 & Quantum theories, technologies, and hardware yet to be matured. Quantum computers are not yet sufficiently fault-tolerant, noise resistant, and have solid programmers and debuggers    & Challenges in integrating quantum computers in 6G due to special requirements, difficulty in performing large scale operations & Difficulties in integrating quantum and classical networks, especially with multiple nodes, premature routing protocols, difficulties in generating redundancies, and communicating over long distances  & Challenges in securing post-quantum networks across wide range of applications, QKD challenges due to noise,  low efficiency of QKD, and using zero-trust with quantum networks         \\ \hline
\textbf{Impact on 6G Adoption}       & Challenges in developing quantum enabled 6G technologies when theories and technology are yet to mature & Challenges to deploy quantum hardware in diverse 6G network landscape     & Challenges in securing connecting quantum computers for network operations and design & Challenges in providing reliable and robust post-quantum communication in 6G 
\\ \hline
\textbf{Proposed Solutions/Current Research Areas}          & PISQ~\cite{bertels2021quantum}, Solid quantum programming languages and debuggers~\cite{metwalli2024testing} & Better quantum error-correction~\cite{krinner2022realizing, sivak2023real}, multi-level wiring for quantum computers~\cite{100quibit}, standardisation~\cite{mazurquantum} & Entanglement routing~\cite{shi2020concurrent}, quantum multiplexers and switches~\cite{yang2023survey, guo2020fidelity}, reliable quantum communication~\cite{hasan2023quantum} & Solid QKD~\cite{cao2022evolution}
, quantum safe classic cryptography~\cite{yang2023survey}, hybrid approaches of QKD and classic cryptography \\ \hline
\textbf{TRL related to 6G Deployment} & TRL 4 – Early-stage                 & TRL 4 – Early-stage     & TRL 4 – Early-stage                     & TRL 4 – Early-stage                      \\ \hline
\end{tabular}
\vspace{-0.5cm}
\end{table*}
\color{black}

\section{Lessons Learned and Future  Research Directions} 
\label{sec:lesson}



This section discusses lessons learned and summarizes future research directions in the categories of \ac{RAN}, core, edge \& transport, security \& privacy and deployment challenges.



\subsection{RAN}

\subsubsection{Lessons Learned}
The research reveals several key insights into the potential of mm-waves and THz frequencies for \ac{QKD} \cite{zhang2023millimeter}: i) Cryogenic Operation: Operating sources and detectors at cryogenic temperatures significantly enhances the performance of \ac{QKD} systems. The study found that lower temperatures, such as T = 4 K, allow for better secret key rates and extended transmission distances. ii) MIMO Advantage: Implementing MIMO technology, as opposed to SISO, can improve secure transmission distances. With an increased number of antennas (e.g., 1024 × 1024), the system can achieve longer secure communication channels. iii) Frequency and Temperature Relationship: There is a notable correlation between frequency, temperature, and secure communication distance. Lower frequencies and cryogenic temperatures contribute to maximising the transmission distance. Several key insights were derived from the research \cite{jang2023cooperative}: \textit{(i) Enhanced Security Through AN Injection:} The introduction of AN injection into the CB-based \ac{PLS} scheme significantly improves the secrecy rate, particularly when the eavesdropper is close to the intended receiver. This method effectively mitigates the vulnerability found in conventional CB-based \ac{PLS} schemes under similar conditions. \textit{(ii) Optimisation of Secrecy Rate:} The study finds that the secrecy rate can be optimised by adjusting the ratio between the data beamformer and AN injection beamformer components. This optimisation is crucial for maintaining high security in varying conditions, including the presence of \ac{CSI} errors. \textit{(iii) Impact of System Parameters:} The research highlights that system parameters, such as the number of \ac{VAA} elements and the size of the \ac{VAA}, have a significant impact on the secrecy rate. Understanding these impacts is essential for designing and operating CB-based \ac{PLS} systems effectively.

\subsubsection{Remaining Research Questions}

Despite significant advancements in \ac{QKD} and \ac{PLS} technologies, several research questions remain to enhance their practical applications and integration with emerging technologies. The following questions outline key areas where further investigation is needed:



\begin{itemize}[leftmargin=*]
    \item \textit{What practical hardware can be developed to reliably operate at cryogenic temperatures and within the targeted frequency range for mm-wave and THz \ac{QKD} systems? What are the most efficient and cost-effective sources and detectors for these conditions?} 
    \item \textit{How scalable are mm-wave and THz \ac{QKD} systems in real-world applications, such as satellite communications and short-distance indoor networks? What challenges must be addressed to implement these systems effectively?}
    \item \textit{How can antenna configurations in MIMO systems be optimized for different environments and distances in mm-wave and THz \ac{QKD} systems? What factors should be considered to enhance their performance across various settings?}
    \item \textit{How can the proposed cooperative beamforming (CB)-based \ac{PLS} with artificial noise (AN) injection be implemented in real-world 6G IoT networks? What insights can be gained from evaluating its practicality and effectiveness under actual network conditions?}
    \item \textit{How can the proposed \ac{PLS} scheme be scaled for large-scale IoT networks with diverse devices and varying resource constraints? What resource management strategies are crucial for its efficient operation?}
    \item \textit{How can the proposed CB-based \ac{PLS} with AN injection be integrated with other existing or emerging security protocols to provide a comprehensive security solution for 6G networks? What are the potential benefits and challenges of such integration?}
    \item \textit{How can the proposed Joint Active Spatial Precoding and Anonymization (JASPA) strategy be adapted for ultra-dense network scenarios with a high concentration of users and eavesdroppers? What modifications would be necessary to maintain security and efficiency in such environments?}
    \item  \textit{What are the potential impacts of integrating the JASPA strategy with ultra-massive MIMO systems, autonomous Internet of Things (IoT), and network slicing technologies? How can these integrations enhance the overall performance and security of the network?}
    \item \textit{How can the \ac{QBSA} be further optimized or modified to cater to heterogeneous networks with diverse user and eavesdropper distributions? What specific factors need to be considered to improve its effectiveness in these varied conditions?}
    
\end{itemize}

\subsubsection{Possible Future Directions}
Future research could focus on several promising directions to enhance mm-wave and THz \ac{QKD} systems \cite{zhang2023millimeter}: 

\begin{itemize}[leftmargin=*]
    \item \textit{ Developing advanced cryogenic cooling systems that are efficient and portable would make \ac{QKD} systems more feasible for widespread use. This includes improving the integration of cryogenic technologies with existing communication infrastructures. } 

 \item \textit{ Investigating alternative materials and technologies that can operate efficiently at high frequencies without requiring extremely low temperatures could reduce the reliance on cryogenic cooling.} 
 
  \item \textit{Combining mm-wave and THz \ac{QKD} systems with other quantum communication technologies, such as satellite-based \ac{QKD} or fibre-optic \ac{QKD}, could create more robust and versatile quantum networks.} 

 \item \textit{Developing adaptive techniques that dynamically adjust the AN injection based on real-time network conditions and the behaviour of potential eavesdroppers could further enhance the security of the proposed scheme \cite{jang2023cooperative}.} 

 \item \textit{Combining the CB-based \ac{PLS} with AN injection with other quantum-resistant cryptographic methods may offer a robust solution against both classical and quantum-based attacks \cite{jang2023cooperative}.} 

 \item \textit{Research into advanced hardware capable of supporting the proposed scheme, especially in terms of antenna design and processing capabilities, will be essential for practical deployment \cite{jang2023cooperative}.} 
\end{itemize}

\subsection{6G Core, Edge and Transport}


\subsubsection{Lessons Learned}

\ac{QC} has shown significant potential for accelerating complex computational tasks that are important both in the network core and at the network edge. For example, quantum algorithms can optimize network routing \cite{alanis2014quantum}, resource allocation \cite{dani2021optimization}, and traffic management \cite{neukart2017traffic} more efficiently than classical approaches, especially in highly dynamic and resource-constrained environments. The integration of \ac{QC} at the network edge enables decentralized processing that reduces latency and improves the responsiveness of real-time applications. This is particularly beneficial for applications such as autonomous vehicles, smart cities and IoT, where fast decision-making is critical \cite{bhatia2024quantum}.  In addition to quantum computing at the core and edge of the network, advances in 6G transport play a crucial role in high-speed, low-latency data transmission between distributed network elements. The introduction of \ac{THz} communications and fiber-based quantum transport systems can increase data transmission speeds, reduce bottlenecks and improve coordination between edge devices and core data centers. Quantum-based optical transport networks promise more efficient management of data loads, especially in dense urban environments where latency is critical for applications such as real-time holography and autonomous systems. Even though \ac{QC} is promising, its integration into existing network infrastructures poses a major challenge. Issues such as the scalability of quantum processors \cite{gambetta2020ibm}, error rates \cite{yang2023survey} and the need for quantum-classical hybrid systems \cite{fan2022hybrid} remain critical barriers to widespread adoption. \ac{QC} at the network edge can contribute to energy savings by reducing the need for data transmission to central cloud servers. However, the energy consumption of the quantum processors themselves is an area that needs to be further optimized to ensure overall energy efficiency.

\textcolor{black}{Quantum estimation and control can also play a fundamental role in the practical implementation of quantum technologies in 6G networks, but have been less explored in the current literature on quantum communication for future wireless systems. Precise quantum parameter estimation is essential for calibration, synchronization and adaptive error correction of quantum communication and computing components. Furthermore, real-time quantum control is crucial for stabilizing fragile quantum states under noisy or dynamic network conditions. While most previous work has focused on algorithmic applications (e.g., quantum key distribution or quantum machine learning), the operational layer — including how quantum devices are tuned, measured, and stabilized — remains a technical bottleneck for real-world deployment. To realize the full potential of quantum-enabled systems in practical 6G scenarios, robust estimation and control is required.}

\subsubsection{Remaining Research Questions}

Despite the progress made, several research questions remain unanswered:




\textcolor{black}{
\begin{itemize}[leftmargin=*]
\item \textit{How can quantum algorithms enhance network optimization in the 6G core?}
 \item \textit{What role does quantum reinforcement learning play in proactive congestion control?}
 \item \textit{What is the impact of quantum computing at the edge for time-sensitive applications like autonomous driving or XR?}
 \item \textit{How can quantum technologies support secure and long-distance communication in 6G transport layers?}
 \item \textit{How can quantum-enhanced signal processing improve data throughput, interference control, and noise reduction in 6G networks?}
 \item \textit{What are the potential benefits of using quantum computing in digital twin simulations and predictive network modeling?}
 \item \textit{How can quantum computing contribute to self-healing and fault-tolerant mechanisms in critical 6G infrastructure?}
 \item \textit{How can quantum computing assist in training and optimizing large-scale AI models and generative applications at the edge?}
 \item \textit{What quantum estimation techniques can be applied for adaptive calibration of distributed quantum devices in 6G networks?}
 \item \textit{What are the trade-offs between speed, fidelity, and energy efficiency in implementing quantum control for network-scale applications?}
\end{itemize}
}

\subsubsection{Possible Future Directions}

\textcolor{black}{As 6G networks evolve to support unprecedented levels of connectivity, intelligence, and complexity, quantum computing emerges as a transformative enabler, offering novel solutions to overcome scalability, optimization, and real-time decision-making challenges across the core, edge, and transport layers.}

\textcolor{black}{
\begin{itemize}[leftmargin=*]
 \item \textit{Develop integrated frameworks that enable quantum algorithms like QAOA and quantum annealing to work alongside classical SDN/NFV controllers for real-time network orchestration. Additionally, integrate quantum control algorithms with network-layer decision-making to enable adaptive routing or encoding based on real-time quantum measurements. This approach will enhance the ability of 6G core, edge and transport  to dynamically optimize routing, slicing, and policy enforcement under strict latency and scalability requirements.} 
 \item \textit{Incorporate quantum-enhanced reinforcement learning (Q-RL) agents into network digital twins to simulate, learn, and adapt to complex traffic dynamics and congestion scenarios. These agents can accelerate policy convergence and improve decision-making for proactive congestion control and mobility management \cite{nawaz2019quantum}.} 
 \item \textit{Integrate small-scale quantum co-processors with GPUs and AI accelerators at the edge to enable hybrid inference for real-time tasks like XR rendering and autonomous control. These architectures can reduce decision latency and improve energy efficiency for edge computing \cite{liu2023enabling}.} 
 \item \textit{Use quantum algorithms to optimize training and fine-tuning of generative models such as LLMs, VAEs, and diffusion models deployed in constrained edge environments. Also investigate generative AI-assisted quantum estimation frameworks that predict drift, decoherence, or error rates across quantum hardware elements. This can lead to faster model adaptation and context-aware personalization with lower computational overhead \cite{dong2022quantum}.} 
 \item \textit{Develop scalable quantum repeater infrastructure using entanglement swapping and error correction to support secure long-distance quantum communications. Quantum teleportation protocols can offer a fundamental shift in secure 6G transport without relying on classical signal replication \cite{azuma2023quantum}.} 
\end{itemize}
}

\subsection{Security and Privacy}

\subsubsection{Lessons Learned}

Quantum technologies, particularly \ac{QKD}, have the potential to greatly enhance the security of 6G networks. \textcolor{black}{They are trustworthy but infrastructure-intensive.} \textcolor{black}{QKD systems require trusted relay nodes, dedicated quantum channels (typically fiber or satellite), and operate over limited distances. This raises scalability and cost-efficiency concerns, especially in heterogeneous or mobile 6G deployments. Trade-offs arise between ideal security and practical deployment feasibility.}  \ac{QSDC} takes this a step further by enabling the direct transmission of information without the need for classical encryption, ensuring that any interception attempts are immediately detected.  \textcolor{black}{While this reduces the complexity of the protocol, its implementation is limited by the current limitations of quantum memory, synchronization and error resilience. It offers the highest security assurance, but is not mature enough for deployment outside tightly controlled testbeds.} Meanwhile, \ac{PQC} is critical to protecting 6G networks from the emerging threats posed by quantum computers, safeguarding classical cryptographic systems from future quantum attacks. \textcolor{black}{However, unlike QKD and QSDC, PQC relies on assumptions about quantum hardness, making it vulnerable to future advances in quantum algorithms. The trade-off is between ease of adoption and long-term cryptographic resilience.}  \textcolor{black}{\ac{QEC} is a prerequisite for fault-tolerant quantum communication, enabling stable transmission over noisy quantum links. Yet, implementing QEC requires significant qubit overhead, highly reliable gates, and extremely low error rates—still a challenge for current NISQ devices. The trade-off lies in reliability versus resource complexity, often limiting real-time applications in current networks.}



\subsubsection{Remaining Research Questions}
Even though there are many research problems have been solved, there are many questions to further investigate. Couple of them are listed below:



\begin{itemize}[leftmargin=*]
    \item \textit{How can \ac{QKD} be effectively scaled to cover larger geographic areas, especially given the limitations of quantum channels in terms of range and noise sensitivity?}
    \item \textit{What are the optimal methods for integrating \ac{QSDC} within 6G networks without introducing excessive hardware complexity or latency?}
    \item \textit{How can PQC be made more computationally efficient to prevent latency and throughput issues in 6G networks, particularly for resource-constrained devices like IoT sensors?}
    \item \textit{What advancements in QEC are required to ensure seamless hybrid quantum-classical error correction in 6G networks, especially for \ac{uRLLC} applications?}
\end{itemize}

\subsubsection{Possible Future Directions}
\color{black}
There are several possible future directions for security and privacy of the 6G networks as listed below:

\begin{itemize}[leftmargin=*]
\item \textit{The development of quantum repeaters and advanced quantum hardware will be essential for extending the range and enhancing the performance of QKD systems within global 6G networks. Hybrid approaches that combine classical and quantum security mechanisms, such as integrating \ac{PQC} and QKD into existing cryptographic frameworks like IPsec and MACsec, will ensure a smooth transition to quantum-secure 6G infrastructures.} 
\item \textit{Developing lightweight quantum transceiver architectures, hybrid classical-quantum network interfaces, and low-latency quantum repeaters tailored for mobile environments. Additionally, protocol-level co-design of QSDC with 6G service requirements—such as URLLC and mMTC—will be essential.} 
\item \textit{Research into lightweight PQC algorithms, optimized for low-power, resource-constrained devices, will be crucial for securing the extensive IoT ecosystems anticipated in 6G.} 
\item \textit{Machine learning techniques may also be explored to improve quantum error correction, enabling predictive error management for real-time quantum communications in 6G networks.}
\end{itemize}

\subsection{Technical challenges} 
\color{black}
\subsubsection{Lessons Learned}
Advents of quantum computing pledges powerful computing capabilities to become a key underpinning technology of emerging 6G and beyond networks~\cite{de2021survey}. However, realising quantum computing, which still remains a maturing technology, as a fully developed technology demands addressing a number of challenges~\cite{yang2023survey}. Developing quantum computing technologies as  a robust technology that is suitable for time and resource critical conditions in 6G and beyond networks is a once of the significant challenges considering the present state of quantum computers. In similar vein, integrating and interfacing quantum computers with 6G network nodes and infrastructure is also a challenge. In addition, connecting multiple quantum nodes in networks also poses a number of challenges, while integrating quantum computing with 6G also raises a number of security challenges. Additionally, the high cost of utilising quantum computers and the requirement of accurate quantum machine learning technologies also need to be addressed to utilise quantum computing as an enabler of 6G and beyond networks.
\subsubsection{Remaining Research Questions}
Following are key research questions that needs to be addressed towards utilising quantum computing as a significant enabler of 6G and beyond networks.



\begin{itemize}[leftmargin=*]
    \item \textit{How to fully understand the concept of quantum computing and related aspects?}
    \item \textit{What are the ways to enhance the fault-tolerance and resistance to environmental impacts of quantum computing?} 
    \item \textit{How to develop quantum computing hardware that is robust and suitable for utilisation in 6G mobile networks across diverse environments?}
    \item \textit{How to overcome issues such as quantum cloning towards performing network operations such as multicasting, broadcasting, and routing?}
    \item \textit{What are the methods of making quantum computing more affordable to perform computation requirements, such as learning and optimisation, in 6G networks?}
\end{itemize}

\subsubsection{Possible Future Directions}
Researchers are actively working towards addressing a wide range of technical challenges related to developing and deploying quantum computing in 6G and beyond networks.
\begin{itemize}[leftmargin=*]
    \item \textit{Developing quantum computing theories to fully understand the concept and develop quantum hardware suitable for diverse and demanding environments~\cite{yang2023survey, 100quibit}.}
    \item \textit{\ac{QKD} is being actively developed to secure quantum networks while exploring the use of classical cryptography and hybrid approaches~\cite{cao2022evolution}.}
    \item \textit{Novel methods to improve the fidelity of quantum teleportation are being explored to increase the robustness of quantum networks across heterogeneous communication channels~\cite{guo2020fidelity, hasan2023quantum}.}
    \item \textit{In addition to the optimization of quantum computers for the use of existing machine learning techniques, quantum machine learning is also being researched in order to solve optimization problems in quantum 6G networks more efficiently~\cite{rajak2023quantum}.}
\end{itemize}


\color{black}

\begin{table*}[h!]
 \captionsetup{font=sc,  position=above, justification=centering, labelsep=newline,singlelinecheck=true}
\centering
\scriptsize

\caption{Summarized Lessons Learned and Future Research Directions for Quantum Technologies in 6G. 
}
\label{tab:LessonsLearnedTech}
\scriptsize
\begin{tabular}{|p{1.1cm}|p{4cm}|p{5.7cm}|p{5.4cm}|}
\hline
\textbf{Subsection} & \textbf{Lessons Learned} & \textbf{Research Questions} & \textbf{Future Directions} \\ \hline

\color{black}

\textbf{6G RAN} 
&
\color{black}

{\begin{tabular}[l]{@{}l@{}} 
-- Cryogenic operation improves  QKD \\ key rates.  \\ -- MIMO enhances secure  transmission  \\ distances. \\ --  Lower frequencies improve \\  communication distance. \\ --  AN injection enhances security.    \\  -- Optimisation of secrecy rate. \\  -- Impact of system parameters.\end{tabular}}

& 

\color{black}
 {\begin{tabular}[l]{@{}l@{}} 
 -- How to develop cost-effective cryogenic hardware \\ for mm-wave and THz \ac{QKD}?  \\ -- How scalable are mm-wave/THz \ac{QKD} systems for \\satellite and  indoor networks?  \\ How can antenna configurations in MIMO systems \\ be optimized for different environments and distances \\ in mm-wave and THz QKD systems?  \\  -- How can the proposed CB-based PLS with AN \\injection be implemented in real-world 6G IoT networks? \\-- How can the proposed PLS scheme be scaled for large\\scale IoT networks with diverse devices and varying \\resource constraints? \\-- How can the proposed CB-based PLS with AN \\injection be integrated with other existing or emerging \\security protocols to provide a comprehensive security \\solution for 6G networks? \end{tabular}}

& 

\color{black}
 {\begin{tabular}[l]{@{}l@{}}  -- Develop portable cryogenic systems for \ac{QKD}.  \\ -- Explore hybrid mm-wave and  THz \ac{QKD} \\systems. \\  -- Investigate alternative materials for high\\frequency operation. \\  -- Develop adaptive AN injection techniques. \\ -- Combine CB-based PLS with AN injection and \\other quantum-resistant cryptographic methods. \\  -- Research advanced hardware for practical \\deployment.  \end{tabular}}

\\ \hline


\textcolor{black}{\textbf{6G Core, Edge, Transport}}
& 
\textcolor{black}{
 {\begin{tabular}[l]{@{}l@{}} -- Quantum algorithms solve 6G  \\ core-layer NP-hardproblems.  \\ -- Integration with classical systems  \\ requires  middleware. \\ -- Quantum optimization supports  \\ complex service-level guarantees. \\ -- Quantum-assisted solutions for  \\  performance and spectrum evaluation. \\ -- Quantum scales better in  \\ combinatorially large networks. \\ \end{tabular}}
 }

& 
\textcolor{black}{
 {\begin{tabular}[l]{@{}l@{}} -- How can quantum algorithms enhance network \\ optimization in the 6G core? \\ -- How can hybrid orchestration be implemented  \\ efficiently in real time?  \\  -- How can quantum computing improve service-aware  \\ policy enforcement? \\  -- How do quantum and classical optimization models  \\ compare in core orchestration?   \\ --  How can quantum methods address massive  \\ user/device connectivity?  \\ \end{tabular}}
}
 
& 
\textcolor{black}{
 {\begin{tabular}[l]{@{}l@{}} -- Embed QAOA and quantum annealing into \\ SDN/NFV for routing and resource  allocation. \\  -- Develop APIs and co-processing architectures  \\ to support quantum-classical workload distribution. \\ -- Design QAOA-based resource slicing for  \\ multiple network tenants under mobility and  \\ QoS constraints. \\  -- Compare latency, scalability, optimality of  \\ classical, quantum, and hybrid approaches. \\   -- Evaluate quantum algorithms under ultra-dense  \\ network simulations involving thousands of \\ concurrent optimization flows.   \end{tabular}}}

\\ \hline

\color{black}
\textbf{Security and Privacy in 6G} 
& 
\color{black}
 {\begin{tabular}[l]{@{}l@{}} --\ac{QKD} enables theoretically \\ unbreakable encryption and detection \\ of eavesdropping. \\ 

-- QSDC allows direct message \\ transmission without classical encryption, \\ with immediate interception detection. \\

-- PQC protects classical systems against \\ quantum attacks, ensuring long-term \\ cryptographic security. \\

-- QEC provides fault tolerance by \\ mitigating noise and errors, ensuring \\ reliable quantum communication in 6G. 
 \end{tabular}}

& 

\color{black}
 {\begin{tabular}[l]{@{}l@{}} 
 -- How to extend QKD over large areas despite range and \\ noise limitations in quantum channels. \\ 

-- How to deploy QSDC in 6G with minimal hardware \\ and  latency overhead. \\ 

-- How to optimize PQC for low-latency,  high-throughput \\ use  in resource-constrained 6G devices (e.g., IoT). \\ 

--What improvements in QEC are needed for reliable \\ hybrid  error correction in latency-critical 6G \\ applications like uRLLC.  
\end{tabular}}

& 

\color{black}
 {\begin{tabular}[l]{@{}l@{}} 
 -- Develop quantum repeaters and hardware  to extend \\ QKD  range and performance; \\ explore hybrid QKD + PQC integration into \\ 
 protocols like IPsec/MACsec. \\ 

-- Design lightweight quantum transceivers, hybrid \\ interfaces, and low-latency repeaters; \\ co-design QSDC with 6G requirements. \\ 

-- Optimize PQC for low-power, \\ resource-constrained 6G IoT devices. \\

-- Use machine learning to improve quantum error \\ correction through predictive error management.  
\end{tabular}}

\\ \hline

\color{black}

\textbf{Technical Challenges} 

&

\color{black}

 {\begin{tabular}[l]{@{}l@{}} --QC theories \& technologies need to \\ mature \\ -- Integrating QC with existing  \\ infrastructure is challenging \\ -- Connecting multiple QC nodes needs \\ solid routing and redundancy generation \\ -- Securing QC enabled networks  needs \\ strong post-quantum cryptography \\ -- QKD needs to be more secure,  \\ efficient,  and noise-tolerant  \end{tabular}}

&

\color{black}

 {\begin{tabular}[l]{@{}l@{}} -- How to explore QC concepts and technologies? \\ -- How to enhance fault-tolerance and noise- tolerance \\ of QC \\ -- How to perform quantum cloning enabling quantum \\ networks to reliably perform routing, multicast, and \\ broadcast \\ -- How to make QC cost-efficient? \end{tabular}}

& 

\color{black}

 {\begin{tabular}[l]{@{}l@{}} -- Developing QC theories and technologies to fully \\ understand  concepts such as quantum entanglement \\ -- Develop QKD to be efficient and  noise-tolerant \\ -- Connect QC nodes with efficient and  robust routing, \\ multicast,  and broadcast  through quantum cloning \\ -- Make QC more cost-efficient enabling  mass-\\production  of quantum hardware \end{tabular}}


\color{black}

\\ \hline

\end{tabular}
\vspace{-0.5cm}
\end{table*}

\section{Conclusion}
\label{sec:conclusion}

\ac{QC} offers the opportunity to optimize network management and improve real-time decision making, which \textcolor{black}{are} crucial for the scalability and efficiency required in 6G networks. In this paper, we have explored the transformative potential of quantum technologies in the context of B5G and 6G networks, focusing on key areas such as quantum computing, quantum machine learning, quantum security and the integration of quantum systems at the core, edge and transport layers.  Although the advantages of quantum technologies for 6G are obvious, there are also significant challenges that stand in the way of widespread deployment. The limitations of current quantum hardware, such as scalability, error rates and energy efficiency, need to be addressed to enable the integration of quantum technologies into existing network architectures. In addition, the development of hybrid quantum-classical systems and quantum-safe protocols is essential to overcome these obstacles and ensure seamless interoperability between classical and quantum systems.   On the road to realizing 6G networks, future research has been proposed that focuses on developing more robust quantum algorithms, improving quantum hardware, and addressing the energy and security issues associated with large-scale deployment.

\section*{Acknowledgement}
This work is partly supported by COST Action CA22104 Beingwise and the Science Foundation Ireland under CONNECT phase 2 (Grant no. 13/RC/2077\_P2) project.

 
\bibliographystyle{IEEEtran}
\bibliography{references.bib}

\vfill

\balance

\end{document}